\documentclass[journal]{IEEEtran}
\ifCLASSINFOpdf
\else
\fi
\usepackage{cite}
\usepackage{amsmath,amssymb,amsfonts}
\usepackage{algorithmic}
\usepackage{algorithm}
\usepackage{graphicx}
\usepackage{textcomp}
\usepackage{xcolor}
\usepackage{cases} 
\usepackage{enumerate}
\usepackage{bm}
\usepackage{subfigure}

\begin{document}
\title{Multi-Agent Deep Reinforcement Learning based Spectrum Allocation for D2D Underlay Communications}

\author{Zheng~Li,
Caili~Guo,~\IEEEmembership{Senior Member,~IEEE}
%        and~Jane~Doe
% <-this % stops a space
\thanks{Corresponding author: Caili Guo (email: guocaili@bupt.edu.cn). This work was supported by the National Natural Science Foundation of China under Grant No. 61571062 and No. 61871047.
	
Z. Li is with the School of Information and Communication Engineering, Beijing University of Posts and Telecommunications, Beijing 100876 China.

C. Guo is with the Beijing Key Laboratory of Network System Architecture and Convergence, School of Information and Communication Engineering, Beijing University of Posts and Telecommunications, Beijing 100876, China.
}}% <-this % stops a space


\maketitle

\begin{abstract}
Device-to-device (D2D) communication underlay cellular networks is a promising technique to improve spectrum efficiency. In this situation, D2D transmission may cause severe interference to both the cellular and other D2D links, which imposes a great technical challenge to spectrum allocation. Existing centralized schemes require global information, which causes a large signaling overhead. While existing distributed schemes requires frequent information exchange among D2D users and cannot achieve global optimization. In this paper, a distributed spectrum allocation framework based on multi-agent deep reinforcement learning is proposed, named multi-agent actor critic (MAAC). MAAC shares global historical states, actions and policies during centralized training, requires no signal interaction during execution and utilizes cooperation among users to further optimize system performance. Moreover, in order to decrease the computing complexity of the training, we further propose the neighbor-agent actor critic (NAAC) based on the neighbor users' historical information for centralized training. The simulation results show that the proposed MAAC and NAAC can effectively reduce the outage probability of cellular links, greatly improve the sum rate of D2D links and converge quickly.

\end{abstract}

\begin{IEEEkeywords}
Device-to-device (D2D) communications, multi-agent deep reinforcement learning, spectrum allocation.
\end{IEEEkeywords}

\IEEEpeerreviewmaketitle

\section{Introduction}
\IEEEPARstart{T}{he} popularity of mobile devices and the growth of multimedia applications have placed a high demand on the data transmission rate of wireless networks. Device-to-device (D2D) communication is regarded as a key technology to improve data transmission rate, reduce latency and energy consumption. It's an important part of future 5G and Internet of Things (IoT). Under the assistance of cellular base station (BS), D2D communication allows two nearby cellular users (CUEs) to form a D2D pair and communicate with each other directly without traversing the BS or core network, thus improving the transmit quality significantly due to short transmission distance \cite{kai2019resource}. D2D underlay communication reuses the spectrum of the cellular network to potentially increase spectral efficiency. However, D2D communication generates interference to the cellular network if the radio resources are not properly allocated \cite{feng2013device,min2011reliability,phunchongharn2013resource}. Thus, it is important to properly allocate radio resources to ensure reliability of cellular communication and increase the capacity in D2D underlay cellular networks.

There have been many resource allocation schemes based on traditional optimization methods in the existing literature. However, in future wireless networks where users are dense and the scene changes rapidly, resource allocation mainly faces two challenges. On the one hand, as the number of users increases, acquiring channel state information (CSI) requires huge signaling overhead and assuming that the BS will have the global network information is unrealistic. On the other hand, resource allocation problems are often modeled as combinatorial optimization problems with nonlinear constraints that are difficult to optimize efficiently by traditional optimization methods. Fortunately, reinforcement learning (RL) has been
shown effective in addressing decision making under uncertainty 
\cite{sutton1998introduction}. RL can learn decision policies from historical data and the hard-to-optimize objective issues can be nicely addressed in a RL framework through designing training rewards such that they correlate with the final objective. Moreover, RL for resource allocation can be designed as a distributed algorithm, where each D2D pair is supported by an autonomous agent, which automatically selects a reasonable spectrum for transmission based on the policy learned by RL. Therefore, we use RL to solve the spectrum allocation problem for D2D underlay communications in this paper.

\subsection{Related Work}
Existing resource allocation methods can be divided into centralized and distributed schemes according to their execution modes. In the centralized schemes\cite{phunchongharn2013resource,wu2018high,kose2018resource,kuang2019energy}, the BS is responsible for allocating resources to the CUEs and D2D pairs, and monitoring information such as signal to interference-plus-noise ratio (SINR), CSI, and interference level of each user in the cell range. With global CSI at the BS, various solutions to the channel and power allocation problems of the D2D tier have been proposed in \cite{phunchongharn2013resource} and the references therein. Graph theory is a useful centralized method to solve this kind of resource allocation problems \cite{tamura1999graph, checco2013learning}. A bipartite graph is designed in \cite{wang2017resource}, where CUEs and D2D pairs are modeled as vertexes and the weight of the bipartite graph is the rate of the associated D2D and cellular links. However, centralized schemes require BS to have global network information. Moreover, the complexity of the centralized schemes increases with the number of users, causing enormous computational pressure on the BS.

In order to reduce the signaling overhead and reduce the computing load of the BS, a series of distributed resource allocation methods are proposed. In a distributed approach, there is no central controller and the D2D pairs opportunistically and autonomously reuse the spectrum of CUEs. It still requires frequent exchange of information between adjacent D2D users and requires the devices to perceive the cellular communications to gather information about channel quality and available resource blocks while the BS is not required to obtain global information and participate in calculation. Distributed schemes can work well to large networks, but require complicated interference avoidance algorithms to ensure high quality cellular communications and reliable D2D communications. Some distributed algorithms are based on game theory \cite{zaki2017distributed, nguyen2016distributed, dominic2018distributed}. Game theory is used to model D2D pairs sharing spectrum resources with CUEs as an auction mechanism in \cite{zaki2017distributed}. A distributed resource allocation has been proposed in \cite{nguyen2016distributed} to guarantee a minimum data rate for CUEs and to maximize the D2D average data rate. Interference is controlled using reuse prices and power control game. Since the channel gain information and prices have to be shared among the D2D pairs, large signaling overhead is incurred. Moreover, this type of method usually requires a lot of iterations to converge.

In addition to game theory, machine learning has been considered as an effective tool in solving different network problems in 5G \cite{jiang2017machine, wang2015artificial}. RL is one of the most powerful tools for policy control and intelligent decision making \cite{sutton1998introduction}, which has been widely adopted in wireless communications \cite{li2014tact,koushik2018intelligent,saleem2017clustering}. Recently, a number of works have applied RL to solve the intelligent resource management and decision making problem in D2D underlay networks \cite{luo2014dynamic,zia2019distributed,yang2019intelligent,ye2018deep1,ye2018deep2,ye2019deep,Autonomous,liang2019spectrum}. A Q-learning based resource allocation has been proposed in \cite{luo2014dynamic}. Resources are shared among D2D and cellular users using Q-learning based strategy to maximize the network throughput. A distributed Q-learning based spectrum allocation scheme has been proposed in \cite{zia2019distributed}, where D2D users learn the wireless environment and select spectrum resources autonomously to maximize their throughput while causing minimum interference to the cellular users. Since Q-learning has low convergence speed and may not always suitable to deal with continuous valued state and action spaces, an efficient transfer actor-critic (AC) RL approach has been proposed in \cite{yang2019intelligent} to address the intelligent resource management problem in a D2D-based Internet of Vehicle (IoV) networks. The above works can only be applied to low-dimensional state-action mapping. Recently, deep learning has also been introduced into resource allocation problems.\cite{8553651} leverages the deep long short-term memory (LSTM) learning technique to make localized prediction of the traffic load at the ultra dense	networks (UDN) base station. In \cite{8546798}, a damped three dimensional (D3D) message-passing algorithm (MPA) based on deep learning for resource allocation in cognitive radio networks has been proposed.  A novel deep learning-based traffic load prediction algorithm to forecast future traffic load and congestion in network has been proposed in \cite{8361420}. With deep learning techniques, reinforcement learning has shown impressive improvement. \cite{8736403} exploits a collaborative learning framework that consists of deep learning in conjunction with reinforcement learning for resource scheduling in network slicing. In \cite{ye2019deep}, a decentralized resource allocation mechanism for vehicle-to-vehicle (V2V) communications based on deep RL has been developed, which can be applied to both unicast and broadcast scenarios. All above works model the policy search process in RL as a Markov decision process (MDP), which is true if different agents (D2D pairs) are independently updating their policies at different times. However, if two or more agents (D2D pairs) are updating their policies at the same time, it becomes a multi-agent environment which appears non-stationary. 

There are some resource allocation studies based on multi-agent RL \cite{Autonomous,7824674,8514983,liang2019spectrum}. In \cite{Autonomous}, the resource allocation problem is modeled as a stochastic non-cooperative game and a Q-learning based algorithm is proposed. This method combines Q-learning and game theory to alleviate the instability of the multi-agent environment. However, it cannot be applied to high-dimensional state-action mapping and its convergence to the Nash equilibrium requires a lot of iterations. A fingerprint-based deep Q-network method has been proposed in \cite{liang2019spectrum}. 
This method is a combination of multi-agent RL and deep learning. By giving all agents a common reward, it mitigates the instability of multi-agent environment but makes each agent fail to achieve the higher individual reward.

\subsection{Contribution}
This paper proposes two distributed spectrum allocation frameworks, multi-agent actor critic (MAAC) and neighbor-agent actor critic (NAAC), which are trained centralizedly and executed distributedly. The frameworks set the respective reward for each agent. By sharing all users' historical states, actions and policies in the centralized training, MAAC can mitigate the instability of multi-agent environment and meanwhile ensure that each agent's policy is updated in the direction of increasing individual reward. Moreover, in order to reduce the computing complexity of the training, NAAC is further proposed to share neighbor users' historical information for centralized training.  Our motivation is to learn from historical information how to make decisions (select spectrum) based on the states observed in real time with the help of deep reinforcement learning. These states include instant channel information observed by UEs, etc. We don't use historical information to make decisions, but we collect historical information for learning the reinforcement learning model. Our frameworks can learn a model with generalization capabilities that can make reasonable decisions based on real-time observed states The two frameworks require no information interaction when they are executed, so they significantly save the signaling overhead. In addition, our methods can transfer complex training processes to the BS and significantly reduce the computing complexity of algorithm execution. 

Part of the work related to NAAC was written as a conference paper \cite{li2019multi} which are published in IEEE Globecom 2019. This paper provides a unified multi-agent deep reinforcement learning framework covering MAAC and NAAC for distributed spectrum allocation. In this  paper, we theoretically deduces the feasibility of the proposed framework based on Markov  game theory. And this paper provides a detailed analysis of how the framework is deployed and the computational complexity and performance overhead of the framework. In addition, more implementation details, experimental results and discussions are provided to better understand  the multi-agent deep reinforcement learning based spectrum management scheme. The main contributions of this paper are summarized as follows:
\begin{itemize}
\item In order to more accurately model state transitions in a multi-agent environment, the D2D communication environment is modeled by Markov game for the first time.
\item A multi-agent deep RL framework, MAAC, is proposed. It shares all users' historical states, actions and policies in the centralized training, which mitigates the issues that the multi-agent environment is unstable and the training is difficult to converge. In addition, it takes into account the cooperation between users and the pursuit of higher individual rewards.
\item We find that the historical information sharing of neighbor users is enough to satisfy the stability of training. Therefore, an enhanced learning framework, NAAC, is proposed. While ensuring the convergence of the training, it reduces the computing complexity and is more suitable for complex and varied communication scenarios.
\end{itemize}

\subsection{Paper Organization}
The rest of this paper is organized as follows. Section
II shows the system model. In Section III, we formulate the D2D communication environment as a partially observable Markov game and adopt the MAAC framework to address it. In Section IV, the NAAC framework with low computational complexity is proposed. The simulation results and analysis are presented in Section V. Finally, Section VI concludes the paper. The key mathematical notations used in our paper are listed in Table \ref{Mathematical}.

\begin{table}[!t]
	\caption{Mathematical Notation}
	\begin{center}
		\begin{tabular}{|c|c|}
			\hline
			\textbf{Notations}& \textbf{Physical interpretation} \\
			\hline
			 $ M, N, K $  & Number of CUEs, D2D pairs and RBs \\
			\hline
			$ P^{b}, P^{d} $ & Power of BS and D2D transmitter \\
			\hline
			$ g^{b,c}_{m} $  & Channel gains from the BS to $ C_{m} $\\
			\hline
			$ g^{t,r}_{n} $ & Channel gains from $ D^{t}_{n} $ to $ D^{r}_{n} $\\
			\hline
			$ g^{t,c}_{n,m} $ & Channel gains from $ D^{t}_{n} $ to $ C_{m} $\\
			\hline
			$ g^{b,r}_{n} $ & Channel gains from the BS to $ D^{r}_{n} $\\
			\hline
	    	$ g^{t,r}_{i,n} $ & Channel gains from $ D^{t}_{i} $ to $ D^{r}_{n} $\\
			\hline
			$ \sigma^{2} $ & The power of AWGN\\
			\hline
			$ \xi^{c}_{m,k} $ & SINR of the received signal at $ C_{m} $ from BS in RB $ k $\\
			\hline
			$ \xi^{d}_{n,k} $ & SINR of the received signal at $ D^{r}_{n} $ from $ D^{t}_{n} $ in RB $ k $ \\
			\hline
			$ \mathcal{R}^{c}_{m,k} $ & Data rates of CUE $ C_{m} $ \\
			\hline
			$ \mathcal{R}^{d}_{n,k} $ & Data rates of D2D pair $ D_{n} $\\
			\hline
			$ W $  &  The bandwidth of each RB\\ 
			\hline
			$ \mathcal{S}, \mathcal{A} $ & State space and action space\\
			\hline
			$ s^{t}, a^{t}, r^{t} $ & State, action and reward at time slot $ t $\\
			\hline
			$ G_{i}^{d,t} $ & The instant channel information of the D2D link \\
			\hline
			$ G_{i}^{c,t} $ & The channel information of the cellular link\\
			\hline
			$ I_{i}^{t-1} $ & The previous interference to the link \\
			\hline
			$ K_{i}^{t-1} $ & The RB selected by the D2D link in the previous time slot\\
			\hline
			$ \xi^{c}_{min} $ & The SINR threshold of the CUE\\
			\hline
			$ \mathcal{R}_{i}^{t} $ & Positive reward \\
			\hline
			$ \mathcal{R}_{i}^{t} $ & Negative reward\\
			\hline
			$ p $ & Transition probability\\
			\hline
			$ \gamma $ &  The reward discount factor \\
			\hline
			$ R^{t}_{i} $ & The sum of discounted future reward\\
			\hline
			$ J_{i} $ & The expected cumulative discounted reward\\
			\hline
			$ \pi $ & The policy of reinforcement learning\\
			\hline
			$ Q $ & The action-value function of reinforcement learning\\
			\hline
			$ \mu $ & Deterministic target policy \\
			\hline
			$ \theta^{\mu}, \theta^{Q} $ & The weight of actor network and critic network\\
			\hline
			$ \mathcal{D} $ & Experience replay buffer\\
			\hline
			$ \tau $ &  ``soft'' update factor\\
			\hline
			$ \lambda $  & The number of neighbor D2D pairs taken in NAAC\\
			\hline
			$ \mathcal{N}_{i}^{nb} $ & A set of neighbor D2D pairs of $ D_{i} $\\
			\hline
			$ \mathbf{s}_{i}^{nb}, \mathbf{a}_{i}^{nb} $ & The states and actions of the neighbors of agent $ i $\\
			\hline
			$ U_{l} $ & The number of neurons the $ l $th layer of the actor network \\
			\hline
			$ V_{h} $ & The number of neurons the $ h $th layer of the critic network\\
			\hline
		\end{tabular}
		\label{Mathematical}
	\end{center}
\end{table}

\section{System Model}
\begin{figure}[!t]
	\centering
	\includegraphics[width=3.5in]{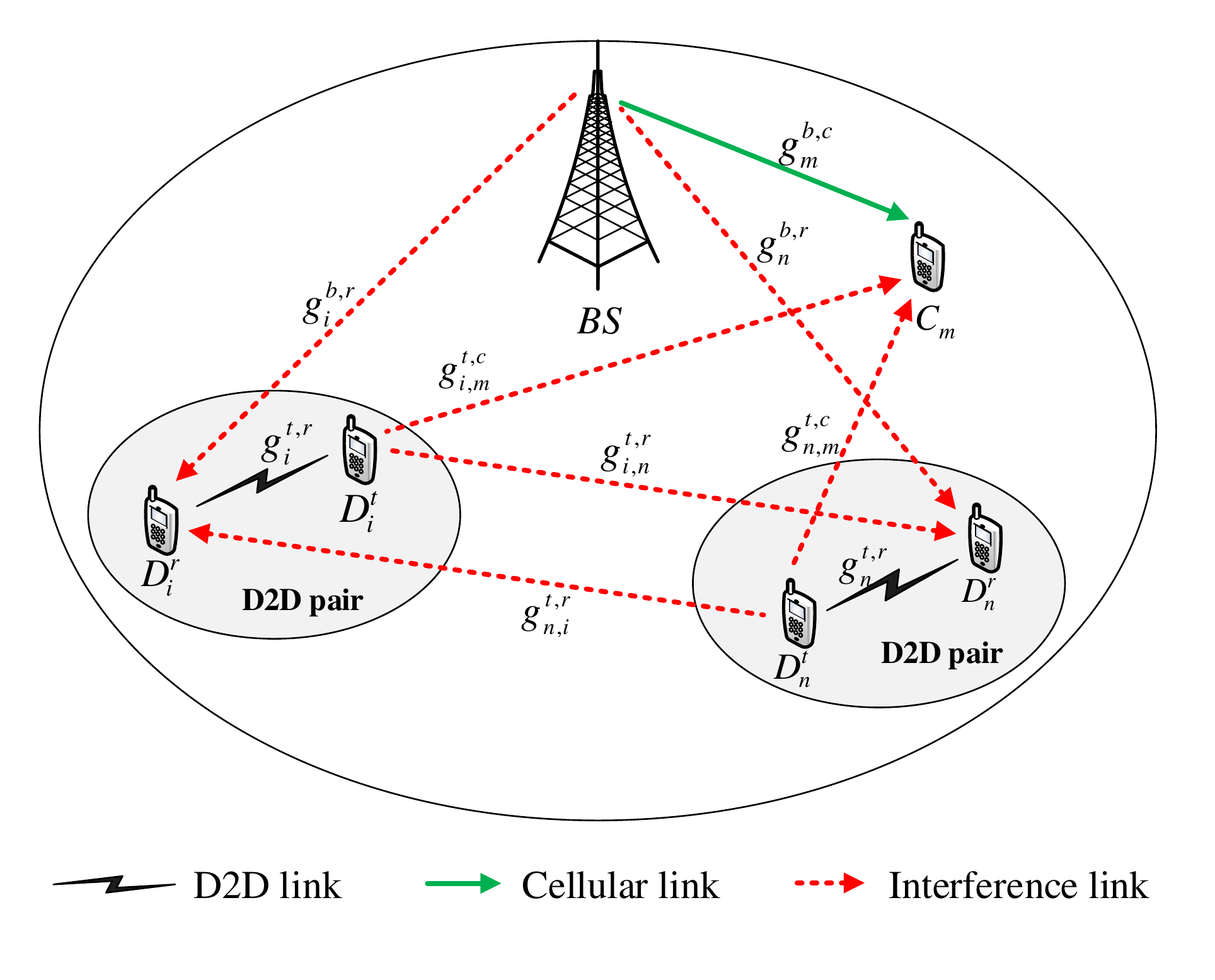}
	\caption{System model for D2D underlaying cellular networks in downlink.}
	\label{D2D_downlink_scenario}
\end{figure}

As illustrated in Fig. \ref{D2D_downlink_scenario}, a downlink scenario in a single cell system is considered. A set of $ M $ CUEs, denoted as $ \mathcal{M} = \{1,...,M\} $, and a set of $ N $ active D2D pairs, denoted as $ \mathcal{N} = \{1,...,N\} $, are located in the coverage area of the base station (BS). We denote the $ m^{th} $ CUE in the system by $ C_{m} $, $ m \in \mathcal{M} $, the $ n^{th} $ D2D pair by $ D_{n} $, $ n \in \mathcal{N} $, the transmitter and the receiver of a D2D pair $ D_{n} $ by $ D^{t}_{n} $ and $ D^{r}_{n} $, respectively. Orthogonal frequency division multiple access (OFDMA) is employed to support multiple access for both the cellular and D2D communications, where a set of $ K $ resource blocks (RBs) are available for spectrum allocation. A RB is the smallest unit of spectrum resources that can be allocated to a user, which is 180 kHz wide in frequency and 1 slot long in time. In this system, the D2D pairs share the same spectrum with the CUEs. There are three types of interference in the system, including:
\begin{itemize}
\item the interference received from  the transmitter of a D2D pair at a CUE;
\item the interference received from the BS at a D2D receiver;
\item the interference received from the transmitter of a D2D pair at the receiver of another D2D pair sharing the same spectrum with that D2D pair.
\end{itemize}

We assume that the BS and the transmitter of a D2D pair transmit with powers $ P^{b} $ and $ P^{d} $, respectively. Denote $ g^{b,c}_{m} $, $ g^{t,r}_{n} $, $ g^{t,c}_{n,m} $, $ g^{b,r}_{n} $, and $ g^{t,r}_{i,n} $ as the channel gains of the cellular communication link from the BS to CUE $ C_{m} $, the D2D communication link from D2D transmitter $ D^{t}_{n} $ to D2D receiver $ D^{r}_{n} $, the interference link from D2D transmitter $ D^{t}_{n} $ to CUE $ C_{m} $, the interference link from the BS to D2D receiver $ D^{r}_{n} $ and the interference link from D2D transmitter $ D^{t}_{i} $ to D2D receiver $ D^{r}_{n} $ when they share the same spectrum for data transmission respectively. The power of the additive white Gaussian noise (AWGN) at a receiver is denoted by $ \sigma^{2} $.

The instantaneous SINR of the received signal at CUE, $ C_{m} $, from the BS in RB $ k $ can be written as
\begin{equation}
\xi^{c}_{m,k}
=\dfrac{P^{b}g^{b,c}_{m}}
{\sum\limits_{n \in \mathbf{D}_{k}}P^{d}g^{t,c}_{n,m}
+\sigma^{2}}, 
\label{cue_sinr}
\end{equation}	
and the instantaneous SINR of the received signal at the D2D receiver, $ D^{r}_{n} $, from the D2D transmitter, $ D^{t}_{n} $, in RB $ k $ can be written as
\begin{equation}
\xi^{d}_{n,k}
=\dfrac{P^{d}g^{t,r}_{n}}
{P^{b}g^{b,r}_{n}
+\sum\limits_{i \in \mathbf{D}_{k},i \ne n}
P^{d}g^{t,r}_{i,n}
+\sigma^{2}},
\label{d2d_sinr}
\end{equation}
where $ \mathbf{D}_{k} $ represents the set of D2D pairs to which RB $ k $ is allocated. 

With the instantaneous SINR, we can find the data rates of CUE, $ C_{m} $, and D2D pair, $ D_{n} $, by
\begin{equation}
\mathcal{R}^{c}_{m,k}
= W \log (1 + \xi^{c}_{m,k}),
\end{equation}
and
\begin{equation}
\mathcal{R}^{d}_{n,k}
= W \log (1 + \xi^{d}_{n,k}),
\end{equation}
where $ W $ is the bandwidth of each RB.

We assume that each CUE has been assigned a RB and a RB can be allocated to multiple D2D pairs, in the mean time, D2D users who need to communicate have already completed pairing before spectrum allocation. When the algorithm is executed, the paired D2D pairs autonomously  selects RBs for communication. Traditionally, resource allocation in D2D communications is formulated as a NP-hard combinatorial optimization problem \cite{plaisted1976some} with non-linear constraints, which is with forbidden complexity. To address this issue, we will investigate multi-agent RL for resource allocation in D2D communications. 

\section{Multi-Agent Deep Reinforcement Learning based Spectrum Allocation} \label{MADRL}
In this section, we first model the multi-agent environment and then a distributed framework based on multi-agent RL is proposed to address the spectrum allocation problem.
\begin{figure*}[!t]
	\centering
	\includegraphics[width=7in]{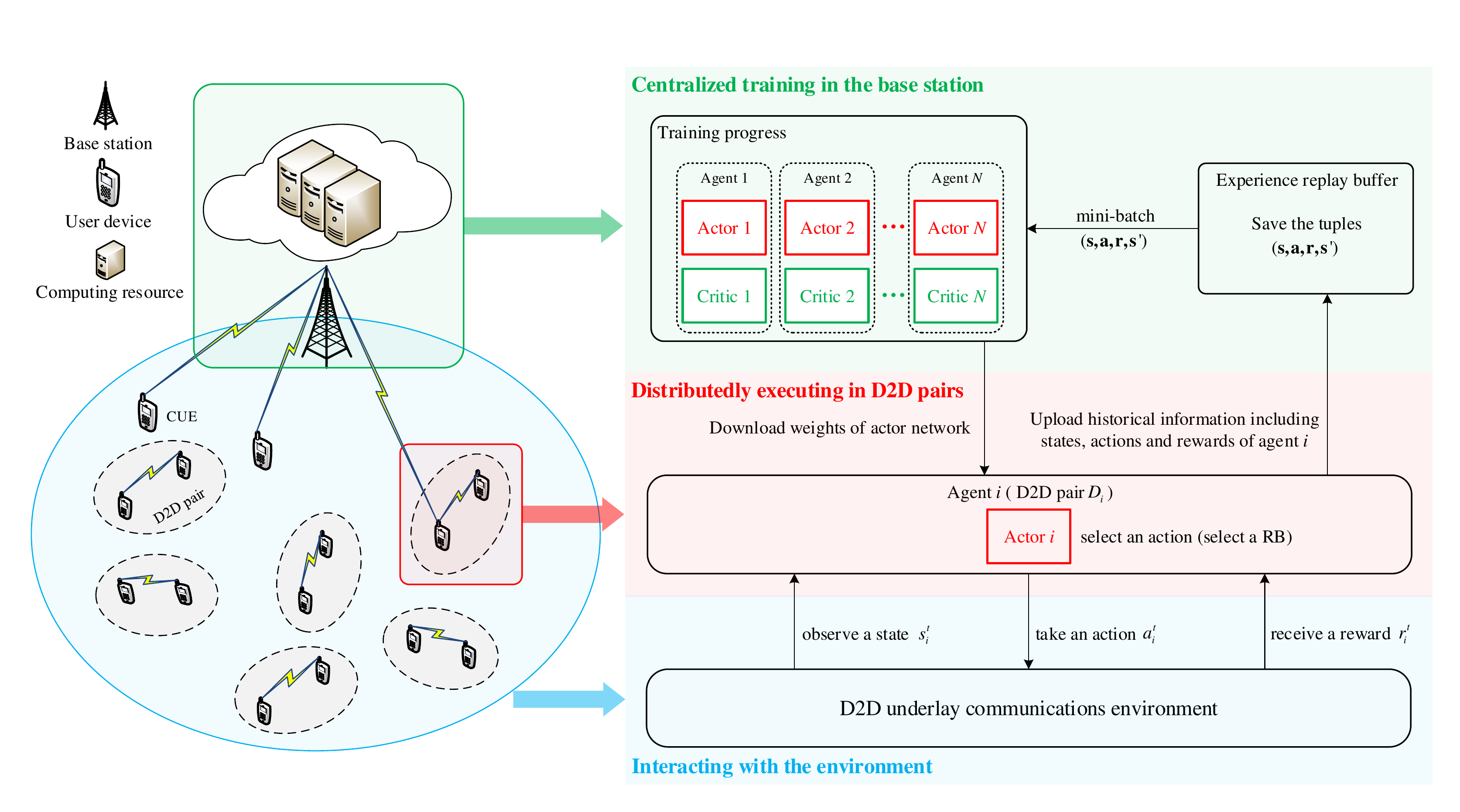}
	\caption{The architecture for MAAC based spectrum allocation in D2D underlay communications.}
	\label{architecture}
\end{figure*}

\subsection{Modeling of Multi-Agent Environments}
In the RL model for D2D underlay communications, an agent, corresponding to a D2D pair, interacts with the environment and takes an action according to a policy. At each time $ t $, the D2D link, as the agent, observes a state, $ s^{t} $, from the state space, $ \mathcal{S} $, and accordingly takes an action (select RBs or power levels), $ a^{t} $, from the action space, $ \mathcal{A} $, based on the policy, $ \pi $. Following the action, the state of the environment transits to a new state $ s^{t+1} $ and the agent receives a reward, $ r^{t} $.

In our system, the state space, $ \mathcal{S} $, the action space, $ \mathcal{A} $, and the reward function, $ r^{t} $, are defined as follows:

\textbf{State space:} The state observed by the D2D link $ D_{i} $ (agent $ i $) for characterizing the environment consists of several parts: the instant channel information of the D2D corresponding link, $ G_{i}^{d,t} $, the channel information of the cellular link, e.g., from the BS to the D2D transmitter, $ G_{i}^{c,t} $, the previous interference to the link, $ I_{i}^{t-1} $, the RB selected by the D2D link in the previous time slot, $ K_{i}^{t-1} $. Hence, $ s_{i}^{t} = [G_{i}^{d,t}, G_{i}^{c,t}, I_{i}^{t-1}, K_{i}^{t-1}] $. The instant channel information and the interference received reveal the quality of each channel.

\textbf{Action space:} At each time $ t $, the agent $ i $ takes an action $ a^{t}_{i} \in \mathcal{A} $, which represents the agent select a RB, according to the current state, $ s^{t}_{i} \in \mathcal{S} $, based on the decision policy $ \pi_{i} $. The dimension of the action space is $ K $ if there are $ K $ RBs. Our methods has good scalability. Discretevalued power and one or more RBs selected by D2D pairs can be modeled as actions, whose dimension is $ \alpha \times \beta$ if there are $ \alpha $ sets of optional RBs and the transmission power is discretized into $ \beta $ levels. Therefore, our algorithm can also solve the resource allocation problem of discrete-valued power control joint spectrum allocation. Note that the action selection of each agent should satisfy the constraint $ \xi^{c}_{m,k} \geq \xi^{c}_{min} $, where $ \xi^{c}_{min} $ represents the SINR threshold of the CUE.

\textbf{Reward function:} The learning process is driven by the reward function in the RL. Each agent makes its decision to maximize its reward with the interactions of the environment. As a result, we will design a reward function for this distributed resource allocation problem as following.

The reward function relates to two parts: the D2D link rate and the SINR constraints of CUE. In our settings, the reward remains positive if the SINR constraints are satisfied; it will be a negative reward, $ r_{neg} < 0 $, otherwise. When the D2D pair $ D_{i} $ (agent $ i $) take an action $ a^{t}_{i} $ at current time slot $ t $, then the D2D pair received a positive reward $ \mathcal{R}_{i}^{t} $, in proportion to the D2D link rate, if the constraints are satisfied. We use the Shannon capacity to evaluate $ \mathcal{R}_{i}^{t} $,
\begin{equation}
\mathcal{R}_{i}^{t} = \log
(1 + \xi^{d,t}_{i}),
\end{equation}
where $ \xi^{d,t}_{i} $ is the instantaneous SINR of the received signal at the D2D receiver $ D^{r}_{i} $ at current time slot $ t $. Therefore, the reward function can be expressed as,
\begin{numcases}{r_{i}^{t} = }
\mathcal{R}_{i}^{t}, & $ \xi^{c}_{m,k} \geq \xi^{c}_{min} $,\\
r_{neg}, & otherwise.
\end{numcases}

Most of the existing works model the policy search process in RL as a Markov decision process (MDP). In MDP, a sequence of resource management decisions of a learning agent by interacting with the wireless communication environment at some discrete time scale can be defined as a tuple $ (\mathcal{S},\mathcal{A},r^{t},p,\gamma) $, where $ p $ is the transition probability $ p(s^{t+1}|s^{t}, a^{t}) $ when the agent takes the action $ a^{t} \in \mathcal{A} $ from the current state $ s^{t} \in \mathcal{S} $ to a new state $ s^{t+1} \in \mathcal{S} $, and $ \gamma \in [0, 1) $ is a discount factor.

However, in the decentralized settings of spectrum allocation problem, all D2D links as agents are independently updating their policies as learning progresses, which is a multi-agent environment if two or more agents updating simultaneously, the environment appears non-stationary from the view of any one agent, violating Markov assumptions required for convergence of RL, and causing instability in the training process. 

To make up for the shortcomings of MDP, we consider a multi-agent extension of MDP in this work called partially observable Markov games, modeling the multi-agent RL. In the multi-agent RL model for D2D underlay communications, at each time $ t $, the D2D link $ D_{i} $, as the agent $ i $, observes a state, $ s^{t}_{i} $ , from the state space, $ \mathcal{S} $, and accordingly takes an action, $ a^{t}_{i} $, from the action space, $ \mathcal{A} $, selecting RB based on the policy $ \pi_{i} $.  Following the action, the state of the environment observed by agent $ i $ transits to a new state $ s_{i}^{t+1} $ and the agent receives a reward, $ r_{i}^{t} $.

An N-agent Markov game is formalized by a tuple $ (\mathcal{S},\mathcal{A},r_{1}^{t},...,r_{N}^{t},p,\gamma) $, where $ \mathcal{S} $ denotes the state space, $ \mathcal{A} $ is the action space, which is assumed to be same for all agents (D2D pairs), $ r_{i}^{t} $ is the reward function for agent $ i $ (D2D pair $ D_{i} $), $ p $ is the transition probability $ p(s_{i}^{t+1}|s_{i}^{t}, a^{t}_{1},...,a^{t}_{N}) $ when all agents take actions $ \{a_{i}^{t} \in \mathcal{A}, i \in \mathcal{N} \} $ simultaneously from the current state $ s_{i}^{t} \in \mathcal{S} $ to a new state $ s_{i}^{t+1} \in \mathcal{S} $. Compared to MDP, Markov game is more accurate in modeling state transitions. The constant $ \gamma \in [0, 1) $ represents the reward discount factor across time. At time step $ t $, all agents take their actions simultaneously, each receives the immediate rewards $ r_{i}^{t} $ as a consequence of taking the previous actions of all agents. The return of agent $ i $ from a state is defined as the sum of discounted future reward
\begin{equation}
R^{t}_{i} = \sum_{n=0}^{T} \gamma^{n} r^{t+n}_{i},
\end{equation}
where $ T $ is the time horizon.

The goal of mulit-agent RL is to learn a policy for each agent to maximize the expected return from the start distribution defined as the expected cumulative discounted reward
\begin{equation}
J_{i} = 
\mathbb{E}[R_{i}^{0}] = \mathbb{E}[\sum\limits_{n=0}^{\infty}\gamma^{n}r_{i}^{n}].
\end{equation}

\subsection{Multi-Agent Actor Critic for Spectrum Allocation}
In order to overcome the inherent non-stationary of the multi-agent environment and to utilize the cooperation between the agents, a multi-agent actor-critic (MAAC) framework is adopted to optimize the policy by modeling multi-agent environment as Markov game and considering action policies of other agents so as to successfully learn policies that require complex multi-agent coordination. In addition, MAAC can make full use of the cooperation among users to further improve the overall performance of the system.

The architecture for MAAC based spectrum allocation in D2D underlay communications is shown in Fig. \ref{architecture}. Each D2D pair, $ D_{i} $, is supported by an autonomous agent $ i $. MAAC is an extension of AC \cite{lillicrap2015continuous} where each agent is divided into two parts: critic and actor. We allow the policies to use the states and actions of all users to ease training. The deep learning training process will cause a lot of computational overhead. Therefore, we transfer the training process to the BS. In order to transfer the complex training process to the BS, our scheme needs D2D users to upload the historical information collected during the execution to the BS. The centralized training process is done at the BS, where critic is augmented with extra information about the policies of other neighbor agents to evaluate the quality of the action. In the distributed execution process, a D2D pair $ D_{i} $ (agent $ i $) downloads the trained weight of the actor from the BS and loads it into its own actor $ i $. The actor $ i $ selects action (RB) $ a^{t}_{i} $ based on the state $ s^{t}_{i} $ observed by the agent $ i $ from the environment. When the agent $ i $ takes the action $ a^{t}_{i} $, the environment returns a reward $ r^{t}_{i} $. When the communication is in good condition, the D2D pair $ D_{i} $ can upload the historical information including $ (s^{t}_{i}, a^{t}_{i}, r^{t}_{i}) $ collected at the execution time to the BS for subsequent training.
\begin{figure}[!t]
	\centering
	\includegraphics[width=3.5in]{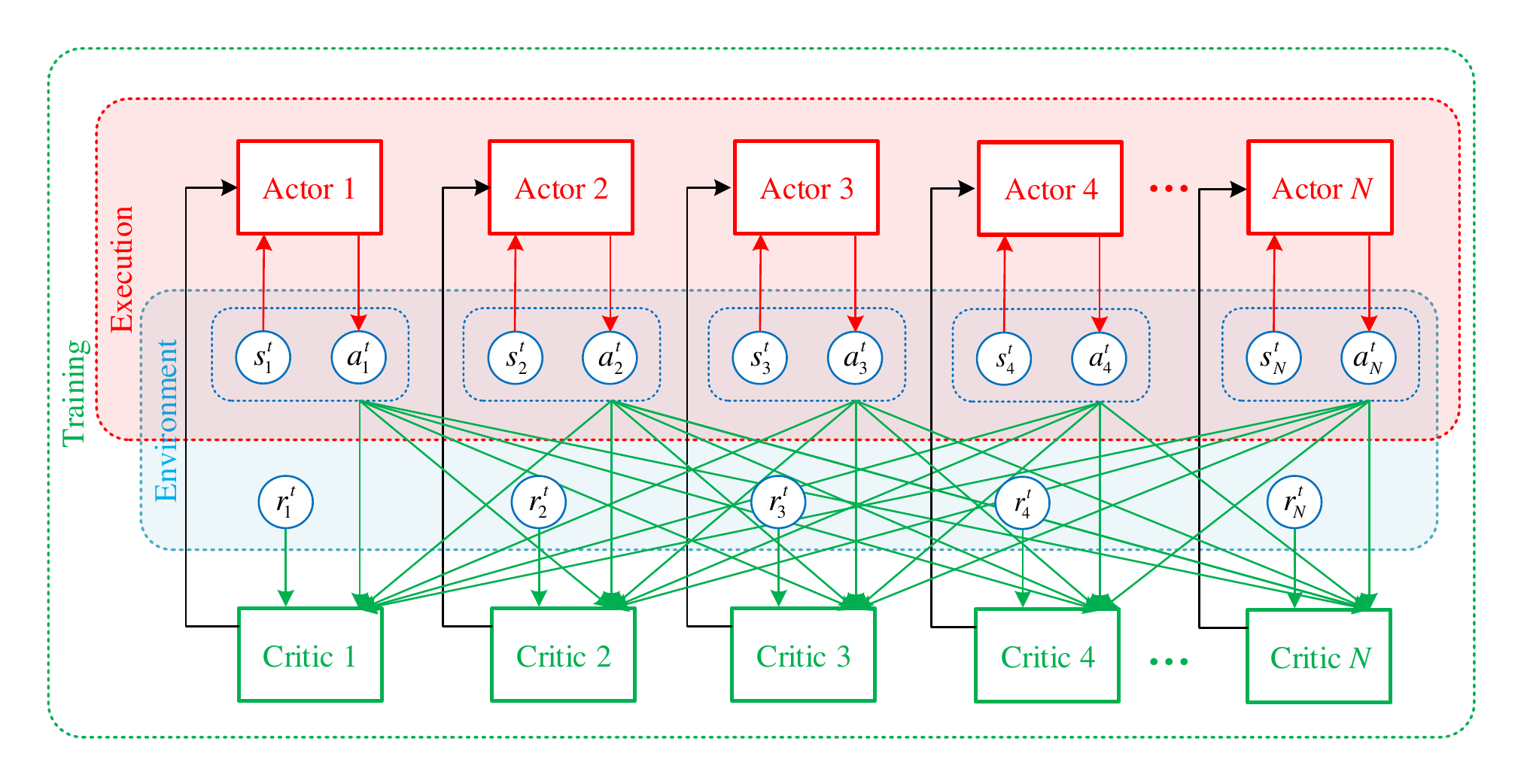}
	\caption{Overview of MAAC of centralized training with decentralized execution.}
	\label{MAAC}
\end{figure}

Overview of MAAC of centralized training with decentralized execution is shown in Fig. \ref{MAAC}. States and actions of all agents are entered into critic to evaluate the quality of the current actions. We allow the policies to use extra information to ease training so long as this information is not used at execution time. It is unnatural to do this with Q-learning based methods, as the Q function generally cannot contain different information at training and test time.

The goal in RL is to learn a policy $ \pi $ which maximizes the expected return from the start distribution $ J = \mathbb{E}_{s \sim E, a \sim \pi}
[R^{0}] $ , where $ E $ denotes the environment. In order to simplify the representation, the state $ s^{t} $, action $ a^{t} $, and return $ R^{t} $ at the current moment are simply denoted as $ s $, $ a $, and $ R $, respectively, $ s^{t+1} $ and $ a^{t+1} $ at the next moment are simply denoted as $ s' $ and $ a' $, respectively. The action-value function is used in many RL algorithms. In a single agent environment, it describes the expected return after taking an action $ a $ in state $ s $ and thereafter following policy $ \pi $:
\begin{equation}
Q^{\pi}(s, a) =
\mathbb{E}_{s,r \sim E, a \sim \pi}
[R].
\end{equation}
Many approaches in RL make use of the recursive relationship known as the Bellman equation \cite{von2007theory}:
\begin{equation}
Q^{\pi}(s, a) =
\mathbb{E}_{s',r \sim E}
\{r + 
\gamma\mathbb{E}_{a' \sim \pi}
[Q^{\pi}(s', a')]\}.
\end{equation}

Extends it into multi-agent environment. Consider a Markov game with $ N $ agents and donete $ \bm{\pi} = \{\pi_{1}, ...,\pi_{N}\} $ as the set of all agent policies. The action-value function (critic) of agent $ i $ can be written as:
\begin{equation}
Q_{i}^{\bm{\pi}}(\mathbf{s}, \mathbf{a}) =
\mathbb{E}_{\mathbf{s}', r_{i} \sim E}
[r_{i} + 
\gamma\mathbb{E}_{\mathbf{a}' \sim \bm{\pi}}
Q_{i}^{\bm{\pi}}(\mathbf{s}', \mathbf{a}')],
\end{equation}
where $ \mathbf{s} $ consists of the states of all agents, $ \mathbf{s} = \{s_{1}, ..., s_{N}\} $, $ \mathbf{a} $ consists of the actions of all agents, $ \mathbf{a} = \{a_{1}, ..., a_{N}\} $, $ Q^{\bm{\pi}}_{i}(\mathbf{s}, \mathbf{a}) $ is a centralized action-value function that takes the states and actions of all agents as input, and outputs the Q-value for agent $ i $. 

If the target policy is deterministic we can describe it as a function $ \mu $ : $ S \leftarrow A $ and avoid the inner
expectation. We now consider $ N $ deterministic policies (actor) denoted as $ \bm{\mu} = \{\mu_{1}, ...,\mu_{N}\} $, The action-value function (critic) of agent $ i $ can be written as:
\begin{equation}
Q_{i}^{\bm{\mu}}(\mathbf{s}, \mathbf{a}) =
\mathbb{E}_{\mathbf{s}', r_{i} \sim E}
[r_{i} + 
\gamma Q_{i}^{\bm{\mu}}(\mathbf{s}', \bm{\mu}(\mathbf{s}'))].
\end{equation}

According to AC \cite{lillicrap2015continuous}, the critic $ Q(s, a) $ can be learned using the Bellman equation as in Q-learning \cite{watkins1992q}. Q-learning is a commonly used off-policy algorithm using the greedy policy $ \mu(s) = \arg \max_{a} Q(s, a) $. We consider the function approximator of the centralized action-value function $ Q_{i} $ of agent $ i $ parameterized by $ \theta_{i}^{Q} $, which we optimize by minimizing the loss:
\begin{equation}
Loss(\theta_{i}^{Q}) =
\mathbb{E}_{\mathbf{s},\mathbf{a},r_{i},\mathbf{s}'}
[(Q_{i}(\mathbf{s}, \mathbf{a}|\theta_{i}^{Q}) - y_{i})^{2}],
\label{loss}
\end{equation}
where
\begin{equation}
y_{i} = r_{i} + \gamma
Q_{i}(\mathbf{s}', \bm{\mu}(s')|\theta_{i}^{Q})
\label{y}.
\end{equation}

Based on the deterministic policy gradient (DPG) algorithm \cite{silver2014deterministic}, a parameterized actor function $ \mu(s|\theta^{\mu}) $ can be used to specify the current policy by deterministically mapping states to a specific action. The policy gradient method is known to exhibit high variance gradient estimates and is exacerbated in multi-agent settings. Since an agent's reward usually depends on the actions (RBs) of many agents (D2D pairs). When the actions of other agents are not considered in the agent's optimization process, the reward conditioned only on the agent's own actions exhibits much more variability, thereby increasing the variance of its gradients.

To analyze the variance of policy gradient methods in multi-agent settings, \cite{lowe2017multi} considers a simple scenario with $ N $ agents and binary actions: $ P(a_{i}=1) = \theta_{i} $. The reward is defined to be $ 1 $ if all actions are the same $ a_{1} = a_{2} = ... = a_{N} $, and $ 0 $ otherwise. Agents must simply learn to either always output $ 1 $ or always output $ 0 $ at each time step. It can prove that the probability of taking a gradient step in the correct direction decreases exponentially with the number of agents, $ N $, which can be expressed in the following proposition.

\textit{Proposition 1:} Consider $ N $ agents with binary actions: $ p(a_{i}=1) = \theta_{i} $, where $ r(a_{1}, ..., a_{N}) = 1 $ when $ a_{1} = a_{2} = ... = a_{N} $, and $ r(a_{1}, ..., a_{N}) = 0 $ otherwise. We assume an uninformed scenario, in which agents cannot get any information from each other and are initialized to $ \theta_{i} = 0.5, \forall i $. Then, if we are estimating the gradient of the $ J $ with policy gradient, we have:

\begin{equation}\label{prop}
p(\langle \hat \nabla J, \nabla J \rangle > 0) \propto (0.5)^{N},
\end{equation}
where $ \hat \nabla J $ is the policy gradient estimator from a single sample, and $ \nabla J $ is the true gradient.

$ P(\langle \hat \nabla J, \nabla J \rangle > 0) $ denotes the probability of taking a gradient step in the right direction that increases reward. Equation (\ref{prop}) indicates that the probability of taking a gradient step in the right direction decreases exponentially, as the number of agents increases.

The high variance gradient estimates of policy gradient methods can be solved by MAAC. The centralized critic in MAAC helps reduce the variance of the gradients since the critic is augmented with extra information about the policies of other agents to remove a source of uncertainty. In addition, conditioned only on the agent's own actions, there is significant variability associated with the actions of other agents, which is largely removed when using these actions as input to the critic.

In MAAC, if the deterministic policy $ \mu_{i} $ of agent $ i $ is parameterized by $ \theta_{i}^{\mu} $, the actor of agent $ i $ is updated by applying the chain rule to the expected return from the start distribution $ J_{i} = \mathbb{E}[R_{i}] $ with respect to the actor parameters:
\begin{equation}
\begin{aligned}
\nabla_{\theta_{i}^{\mu}} J_{i}
& \approx \mathbb{E}_{\mathbf{s}, \mathbf{a} \sim \mathcal{D}}
[\nabla_{\theta_{i}^{\mu}}
Q_{i}(\mathbf{s}, \mathbf{a}|\theta_{i}^{Q})|
_{a_{i} = \mu(s_{i}|\theta_{i}^{\mu})}] \\
& = \mathbb{E}_{\mathbf{s}, \mathbf{a} \sim \mathcal{D}}
[\nabla_{a_{i}}
Q_{i}(\mathbf{s}, \mathbf{a}|\theta_{i}^{Q})|_{a_{i} = \mu_{i}(s_{i})}]
\nabla_{\theta_{i}^{\mu}}\mu_{i}(s_{i}|\theta_{i}^{\mu}).
\end{aligned}
\label{gradient}
\end{equation}
Here $ \mathcal{D} $ is the experience replay buffer contains the tuples $ (\mathbf{s}, \mathbf{a}, \mathbf{r}, \mathbf{s}') $, recording experiences of all agents.

MAAC controls the update of historical state by setting a fixed size experience replay buffer. The experience replay buffer is a finite sized cache. Transitions are sampled from the environment and the tuple $ (\mathbf{s}, \mathbf{a}, \mathbf{r}, \mathbf{s}') $ is stored in the replay buffer. When the replay buffer is full, the oldest samples are discarded. At each timestep the actor and critic are updated by sampling a minibatch uniformly from the buffer. Since MAAC is an off-policy algorithm, the replay buffer can be large, allowing the algorithm to benefit from learning across a set of uncorrelated transitions.

A primary motivation behind MAAC is that, if we know the actions taken by all agents, the environment is stationary even as the policies change \cite{lowe2017multi} since 
\begin{equation}
\begin{aligned}
&p(s_{i}'|s_{i},a_{1},...,a_{N},\pi_{1}, ...,\pi_{N}) \\
&=p(s_{i}'|s_{i},a_{1},...,a_{N}) \\
&=p(s_{i}'|s_{i},a_{1},...,a_{N},\pi_{1}', ...,\pi_{N}')
\end{aligned}
\label{probability}
\end{equation}
for any $ \bm{\pi} \neq \bm{\pi}' $. We use a N-agent Markov game to model the multi-agent RL for D2D underlay communications, where the transition probability is   $ p(s_{i}^{t+1}|s_{i}^{t}, a^{t}_{1},...,a^{t}_{N}) $. So if we know the actions taken by all agents, \ref{probability} is clearly established. The constant transition probability satisfies the Markov assumption of RL convergence. Therefore, the experience replay buffer can be used in MAAC, at the same time the training process of MAAC can mitigate the inherent non-stationary of the multi-agent environment and converge very well. Moreover, the critic considers the actions of all agents to evaluate the quality of the selected action, and can fully utilize the cooperation between the agents.

The mapping between the state space and the action space of the actor part and the action-value function of the critic part need to be approximated by a function approximator. Q-learning works well and a look-up table can be used to accomplish the update rule if the state and action spaces of the problem are small. However, if the state-action space is too large, many states may be rarely visited and thus the corresponding Q-values are seldom updated, leading to a much longer time to converge \cite{ye2019deep}. To solve this problem, deep neural networks (DNNs) are used to approximate the mapping in high-dimensional space. The weight $ \bm{\theta} $ of a DNN is updated by training. Once $ \bm{\theta} $ is determined, a state will correspond to a unique action. The DNN can approximate a complex mapping between high-dimensional spaces based on a large amount of training data that will be used to update $ \bm{\theta} $. 
\begin{figure}[!t]
\centering
\includegraphics[width=3.5in]{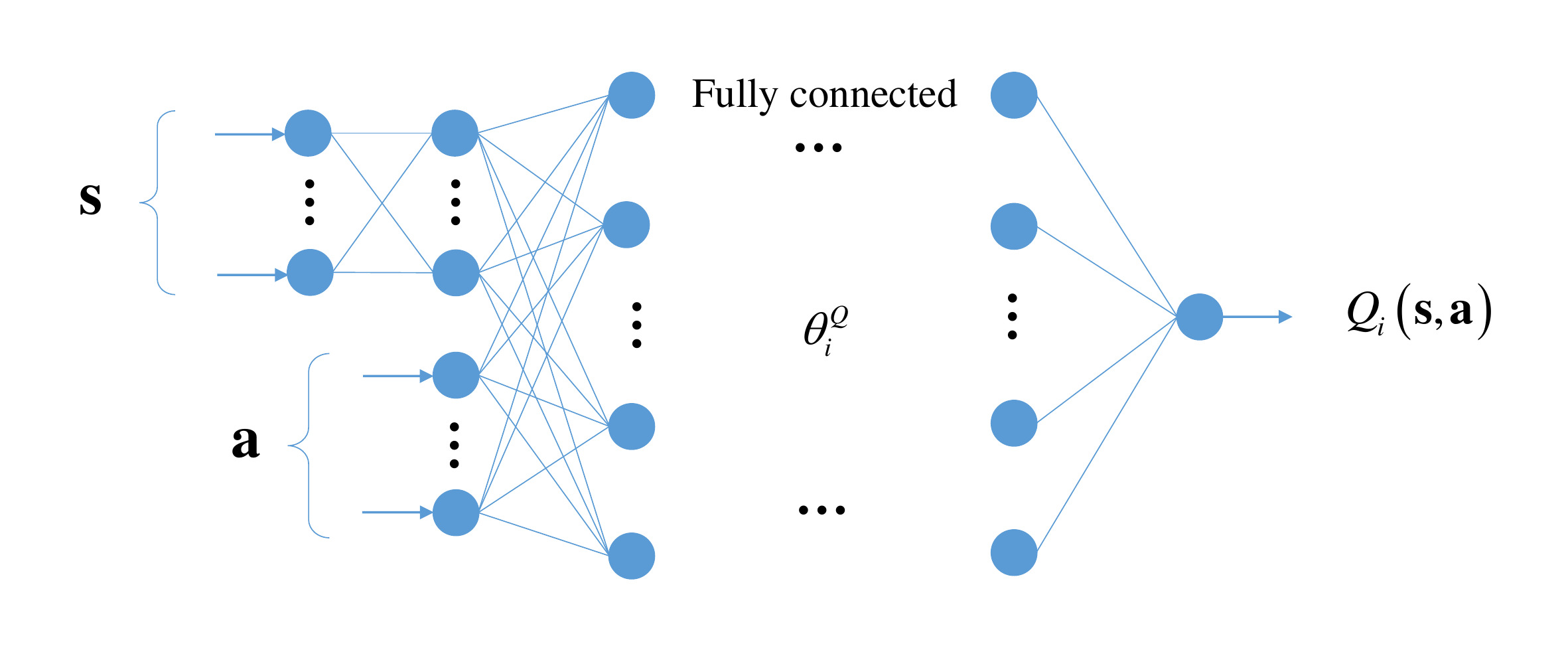}
\caption{Structure of the critic network in MAAC.}
\label{critic}
\end{figure}

In MAAC, we denote the set of actor networks and critic networks of all agents as $ \bm{\mu} = \{\mu_{1}, ...,\mu_{N}\} $ and $ \mathbf{Q} = \{Q_{1}, ...,Q_{N}\} $ with the weights $ \bm{\theta}^{\bm{\mu}} = \{\theta_{1}^{\mu}, ...,\theta_{N}^{\mu}\} $ and $ \bm{\theta}^{\mathbf{Q}} = \{\theta_{1}^{Q}, ...,\theta_{N}^{Q}\} $, respectively. The input of the actor network is the state observed by the agent, and the output is the selected action. The hidden layers in the actor network are all fully connected layers. Fig. \ref{critic} provides the structure of the critic network. The critic network first enters the states of all agents and then a fully connected layer, the actions of all agents then go through several fully connected layers and finally output Q-value. 

Directly implementing Q-learning in equation (\ref{loss}) with neural networks has proved to be unstable in many environments. Since the network $ Q_{i}(\mathbf{s}, \mathbf{a}|\theta_{i}^{Q}) $ being updated is also used in calculating the target value in equation (\ref{y}), the Q-value update is prone to divergence. To solve this problem, we use ``soft'' target updates. We create a copy of the actor and critic networks for every agent, $ \mu'_{i}(s_{i}|\theta_{i}^{\mu'}) $ and $ Q'_{i}(\mathbf{s}, \mathbf{a}|\theta_{i}^{Q'}) $ respectively, that are used for calculating the target values. The weights of these target networks are then updated by having them slowly track the learned networks
\begin{equation}
\theta' \leftarrow \tau \theta + (1-\tau) \theta',
\end{equation}
with ``soft'' update factor $ \tau \ll 1 $. In the experiment, $ \tau = 0.01 $. This means that the target values are constrained to change slowly, greatly improving the stability of learning.

\subsection{Training and Execution}
MAAC framework is divided into training and execution in use. Since the BS has more computing power than the mobile device, the training part of the algorithm is completed at the BS and the users only need to download the weights of the trained target actor network  $ \bm{\theta}^{\bm{\mu}'} $ from the BS, and only uses the actor part to execute the algorithm distributedly. The training algorithm is shown in Algorithm 1. The MAAC framework uses historical information to train the DNNs of the actor part and the critic part, and returns the weights of target actor network $ \bm{\theta}^{\bm{\mu}'} $. New data is generated when the algorithm is executed, which can be added to the experience replay buffer $ \mathcal{D} $ to further fine-tune weights.

The execution algorithm is as shown in Algorithm 2. Users download $ \bm{\theta}^{\bm{\mu}'} $ from the base station and import the weights into their actor networks. All agents input the observed states into the actor networks and the selected actions are output, where the actions correspond to the selected RB.

\section{Neighbor-Agent Deep Reinforcement Learning based Spectrum Allocation}
The MAAC framework proposed in section \ref{MADRL} can make full use of the cooperation relationship among users to improve system performance and at the same time has a high convergence speed. However, MAAC requires information of all agents to assist training. It will result in high computational complexity and high computational overhead, in the case of a large number of users. Therefore, we further reduce the complexity of MAAC without losing too much performance in this section.
\begin{algorithm}
\caption{Training Algorithm}  
\label{Training}
\textbf{Input:} Actor network structure, critic network structure. \\
\textbf{Output:} The weights of target actor network $ \bm{\theta}^{\bm{\mu}'} $. \\
\textbf{Training:}
\begin{algorithmic}[1]
\STATE {Random initialize actor network $ \bm{\mu} $ and critic network $ \mathbf{Q} $ with the weights $ \bm{\theta}^{\bm{\mu}} $ and $ \bm{\theta}^{\mathbf{Q}} $.}
\STATE {Initialize the target actor network $ \bm{\mu}' $ and target critic network $ \mathbf{Q}' $ with weights 
	$ \bm{\theta}^{\bm{\mu}'} \leftarrow \bm{\theta}^{\bm{\mu}} $, 
	$ \bm{\theta}^{\mathbf{Q}'} \leftarrow \bm{\theta}^{\mathbf{Q}} $.}
\STATE {Initialize the experience replay buffer $ \mathcal{D} $}
\STATE {All agents (D2D pairs) receive initial observation states $ \mathbf{s}^{0} = \{s_{1}^{0}, ..., s_{N}^{0}\} $} 
\FOR{each time slot $ t=0,1,...T $}
\STATE {All agents select actions 
	$ \mathbf{a}^{t} = \{a_{i}^{t} = \mu_{i}(s_{i}^{t}), i \in \mathcal{N}\} $ according to the current policy.}
\STATE {All agents execute actions $ \mathbf{a}^{t} $, observe rewards $ \mathbf{r}^{t} = \{r_{1}^{t}, ..., r_{N}^{t}\} $ and observe new states $ \mathbf{s}^{t+1} $.}
\STATE {Save the tuples $ (\mathbf{s}^{t}, \mathbf{a}^{t}, \mathbf{r}^{t}, \mathbf{s}^{t+1}) $ in $ \mathcal{D} $} 
\STATE {Sample a random mini-batch of tuples $ (\mathbf{s}, \mathbf{a}, \mathbf{r}, \mathbf{s}') $ from the $ \mathcal{D} $.}
\STATE {Set $ y_{i} = r_{i} + \gamma
	Q_{i}'(\mathbf{s}', \bm{\mu}'(s'|\bm{\theta}^{\bm{\mu}'})|\theta_{i}^{Q'}) $.}
\STATE {Update critic by minimizing the loss in equation (\ref{loss}).}
\STATE {Update the actor policy using the sampled policy gradient according to equation (\ref{gradient}).}
\STATE {Update the target networks: \\
	$ \bm{\theta}^{\bm{\mu}'} \leftarrow \tau \bm{\theta}^{\bm{\mu}}
	+ (1-\tau) \bm{\theta}^{\bm{\mu}'} $, \\
	$ \bm{\theta}^{\mathbf{Q}'} \leftarrow \tau \bm{\theta}^{\mathbf{Q}} 
	+ (1-\tau) \bm{\theta}^{\mathbf{Q}'} $.}
\ENDFOR
\end{algorithmic} 
\textbf{Return:} $ \bm{\theta}^{\bm{\mu}'} $
\end{algorithm}

\begin{algorithm}
	\caption{Execution Algorithm}  
	\label{Execution}
	\textbf{Input:} Actor network structure, the weights of target actor network $ \bm{\theta}^{\bm{\mu}'} $. \\
	\textbf{Execution:}
	\begin{algorithmic}[1]
		\STATE {Load $ \bm{\theta}^{\bm{\mu}'} $ to the actor network.}
		\STATE {All agents (D2D pairs) receive initial observation states $ \mathbf{s}^{0} = \{s_{1}^{0}, ..., s_{N}^{0}\} $}
		\FOR{each time slot $ t=0,1,...T $}
		\STATE {All agents select actions $ \mathbf{a}^{t} = \{a_{i}^{t} = \mu_{i}'(s_{i}^{t}), i \in \mathcal{N}\} $ according to the policy.}
		\STATE {All agents execute actions $ \mathbf{a}^{t} $ and observe new states $ \mathbf{s}^{t+1} $.}
		\ENDFOR
	\end{algorithmic} 
\end{algorithm}

\subsection{Neighbor-Agent Actor Critic for Spectrum Allocation}
\begin{figure}[!t]
	\centering
	\includegraphics[width=3.5in]{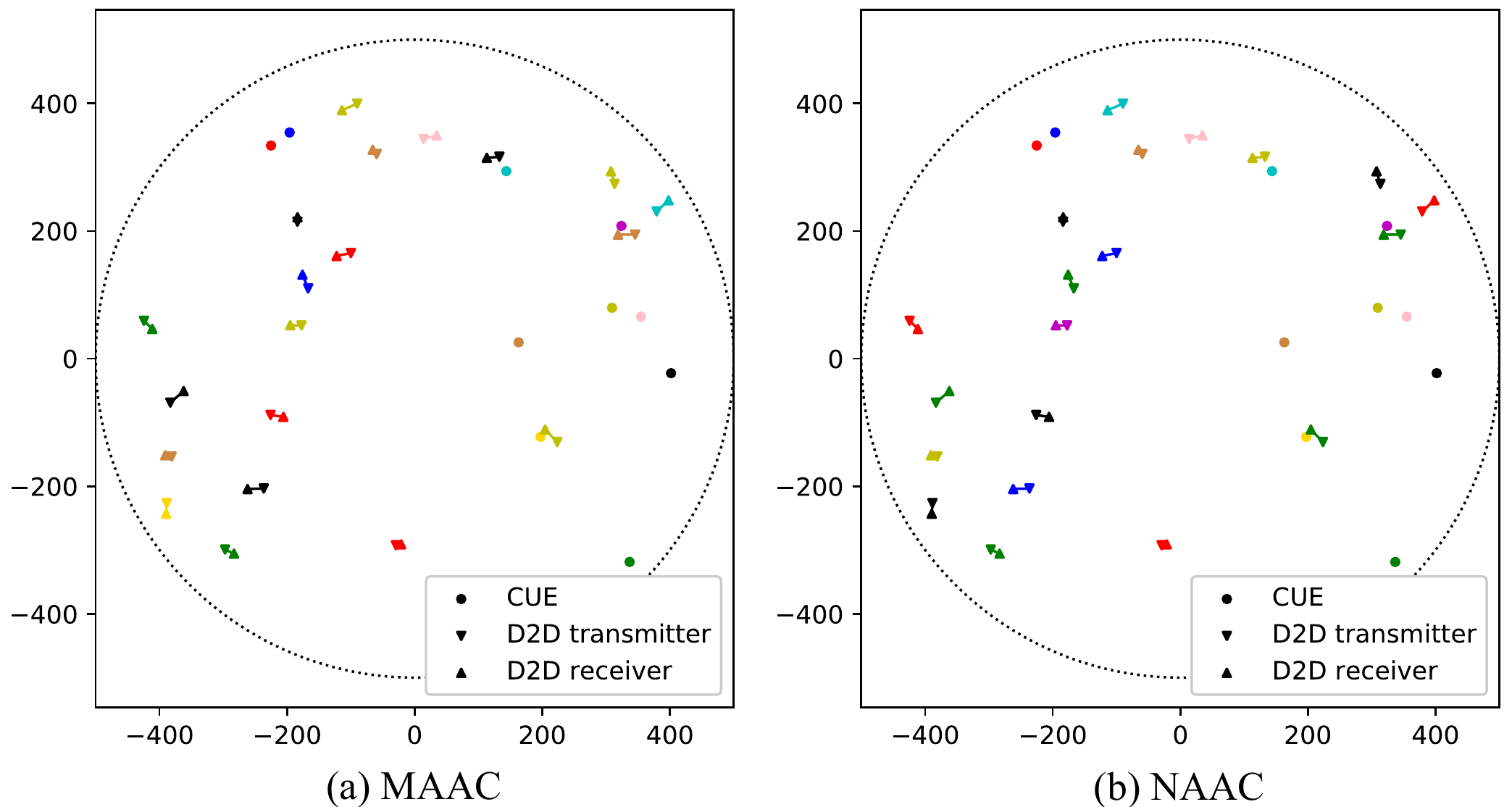}
	\caption{User geographic location and spectrum allocation results visualization of MAAC (a) and NAAC (b).}
	\label{topology}
\end{figure}

In order to ensure the stability of the multi-agent environment and make full use of the cooperation between users, MAAC adds states and actions of all agents to the critic network for training. The geographic location of all CUEs and D2D pairs and the spectrum allocation results after MAAC training convergence are visualized, as shown in Fig. \ref{topology} (a), where different colors represent different RBs and the number of CUEs, D2D pairs and RBs are $ 10 $, $ 20 $ and $ 10 $, respectively. We can see that users who are closer together are allocated different RBs and users who are far apart may share RBs. This is because in a wireless communication environment, inter-user interference is mainly related to the neighbor users. When the user's transmit power is constant, the main factor affecting the inter-user interference strength is the large-scale fading, which is mainly related to the distance between users. Therefore, it is not necessary to have all users' information to ensure the stability of the environment, just the information of the neighbor users is enough. Therefore, we improve MAAC by proposing an improved framework that only allows the states and actions of a fixed number of agents adjacent to the target agent to be added to the critic network for training, called neighbor-agent actor critic (NAAC).

We use distance to define the neighbor users. In NAAC, for a D2D pair $ D_{i} $, we define $ \lambda $ D2D pairs closest to $ D_{i} $ as the neighbor users of $ D_{i} $. We denote a set of D2D pairs which contains $ D_{i} $ itself and its neighbor users as $ \mathcal{N}_{i}^{nb} $. We can use the information of neighbor agents (D2D pairs) instead of global information to ensure the stability of the multi-agent environment since
\begin{equation}
p(s_{i}'|s_{i},\mathbf{a})
\approx p(s_{i}'|s_{i},\mathbf{a}_{i}^{nb}),
\end{equation}
where $ \mathbf{a}_{i}^{nb} = \{a_{j}, j \in \mathcal{N}_{i}^{nb} \} $ contains the actions of the neighbors of agent $ i $.

The loss function and gradient update function of NAAC have the same form as MAAC, the difference is that $ \mathbf{s} $ and $ \mathbf{a} $ in equation (\ref{loss}) and (\ref{gradient}) is changed to $ \mathbf{s}_{i}^{nb} $ and $ \mathbf{a}_{i}^{nb} $, where $ \mathbf{s}_{i}^{nb} $ contains the states of the neighbors of agent $ i $, $ \mathbf{s}_{i}^{nb} = \{ s_{j}, j \in \mathcal{N}_{i}^{nb} \} $.

The actor network in the NAAC framework is identical to that in the MAAC framework. The input of the critic network in the NAAC framework is changed to the states and actions of the neighbor agents. Like MAAC, NAAC uses an experience replay mechanism to overcome the correlation and non-stationary distribution of empirical data. In addition, NAAC also uses ``soft'' target updates to ensure the stability of training.

\subsection{Training and Execution}
The NAAC framework can also be divided into two parts: the training process and the execution process. The execution algorithm of NAAC is exactly the same as MAAC. In the training algorithm, for each time slot $ t $, save the tuples $ (\mathbf{s}^{t}, \mathbf{a}^{t}, \mathbf{r}^{t}, \mathbf{s}^{t+1}) $ in experience replay buffer $ \mathcal{D} $ and sample a random mini-batch of tuples $ (\mathbf{s}, \mathbf{a}, \mathbf{r}, \mathbf{s}') $ from the $ \mathcal{D} $, then set
\begin{equation}
y_{i} = r_{i} + \gamma
Q_{i}'(\mathbf{s}_{i}^{nb'}, \bm{\mu'}(s'|\bm{\theta}^{\bm{\mu}'})|\theta_{i}^{Q'}),
\end{equation}
and update critic by minimizing the loss in equation (\ref{loss}), where $ \mathbf{s} $ and $ \mathbf{a} $ is changed to $ \mathbf{s}_{i}^{nb} $ and $ \mathbf{a}_{i}^{nb} $. Finally, update the actor policy using the sampled policy gradient according to equation (\ref{gradient}), where $ \mathbf{s} $ and $ \mathbf{a} $ is changed to $ \mathbf{s}_{i}^{nb} $ and $ \mathbf{a}_{i}^{nb} $. The remaining steps in the training algorithm are the same as MAAC.

The spectrum allocation results after NAAC training convergence are visualized, as shown in Fig. \ref{topology} (b), where NAAC works with the number of neighbor users $ \lambda=3 $. We find that the historical information sharing of neighbor users is enough to satisfy the stability of training and get a reasonable spectrum allocation result.

\subsection{Computational Complexity and Overhead Analysis}

The computational complexity is critical to the utility of an algorithm. Therefore, we analyze the computational complexity of the two proposed methods at execution time. Define the number of neurons the $ l $th layer of the actor network as $ U_{l} $. The computational complexity of the $ l $th layer is $ O(U_{l-1}U_{l} + U_{l}U_{l+1}) $. The computational complexity of the actor network is $ O(\sum_{l=2}^{L-1} (U_{l-1}U_{l} + U_{l}U_{l+1})) $, where $ L $ is the number of layers of the actor network. The critic network is also a fully connected network. Define the number of neurons the $ h $th layer of the critic network as $ V_{h} $. The computational complexity of the $ h $th layer is $ O(V_{h-1}V_{h} + V_{h}V_{h+1}) $. The computational complexity of the critic network is $ O(\sum_{h=2}^{H-1} (V_{h-1}V_{h} + V_{h}V_{h+1})) $, where $ H $ is the number of layers of the critic network. 
	
For MAAC and NAAC, only the actor network is used during execution, and the actor network they use is the same, so MAAC and NAAC are executed with the same complexity which is $ O(\sum_{l=2}^{L-1} (U_{l-1}U_{l} + U_{l}U_{l+1})) $. Both the actor network and the critic network participate in the training process, so the computational complexity of the training process is $ O(\sum_{l=2}^{L-1} (U_{l-1}U_{l} + U_{l}U_{l+1}) + \sum_{h=2}^{H-1} (V_{h-1}V_{h} + V_{h}V_{h+1})) $. Since MAAC needs to input the states and actions of all users into the critic network during training, the number of neurons in the first layer of the cirtic network of MAAC is more than that of NAAC, so the computational complexity of training is also higher than that of NAAC.

The overhead of system deployment is also important for the utility of the system. The deep learning training  process will cause a lot of computational overhead. It is unrealistic to complete the training on  the mobile device. Therefore, we transfer the training process to the BS, because the BS can  easily deploy hardware devices such as GPUs, it has relatively more computing power. In order to transfer the training process to the BS, our scheme needs D2D users to upload  the historical information collected during the execution to the BS. This historical information  includes the states observed by the D2D users, the actions taken and the rewards they obtained,  all of which are numeric data. The history information generated by a UE in a time slot is only  a few kilobytes in size, which results in small transmission overhead. After the training  process is completed, the device only needs to download the weight of the trained actor network from the BS and import its own actor network to perform spectrum selection. The weight of the neural network is also numerical data. The weight of each UE's actor network is about 300 KB in size, which does not cause too much transmission overhead. In summary, transferring the complex training processes to the BS requires only a small amount of transmission overhead. Traditional strategies require users to report channel status information or exchange information between users in real time, which can cause serious signaling overhead. Our method does  not require users real-time reporting and exchanging information. Our method only requires  the user to upload samples of historical information they have collected to the BS when the  communication is in good condition, which can save a lot of real-time signaling overhead.

\subsection{Comparison of MAAC and NAAC}
Since the input of the critic network in NAAC is a part of the information of the agents, the data dimension is smaller and the required neural network is also smaller, which reduces the computing complexity of the algorithm and improves the training speed. In addition, since the input of the critic network in the NAAC is the states and actions of a fixed number of neighbor users of the target agent, the network structure of the NAAC does not change when the number of D2D pairs in the cell changes. The previously trained weights can continue to be used to speed up the training process. Compared with MAAC, since NAAC does not share all users' states, actions and policies, its modeling of user state transition is not as accurate as MAAC, which will inevitably cause some loss to the convergence of training and the reliability of user communication. However, NAAC has better generalization ability, can be scale well to a larger network, adapt to more varied environments and save computing resource.

\section{Performance Evaluation}
In this section, we compare the MAAC and NAAC with other four distributed approaches:
\begin{itemize}
	\item The most classic RL method Q-learning \cite{watkins1992q}, which is used in \cite{zia2019distributed};
	\item A RL method with better convergence performance, Actor-Critic (denoted as AC) \cite{lillicrap2015continuous}, which is used in \cite{yang2019intelligent} and can deal with continuous valued state and action spacec;
	\item The most classic deep RL method Deep Q Network (denoted as DQN) \cite{mnih2015human}, which uses DNN to approximate the mapping in high-dimensional space \cite{ye2019deep};
	\item A game theory approach, Uncoupled Stochastic Learning Algorithm (denoted as SLA), which is developed in \cite{dominic2018distributed}.
\end{itemize}
Since we assume that each D2D pair can only obtain its own CSI and there is no information exchange among D2D users, centralized approaches with global information do not participate in performance comparisons.

For the simulation, we consider a single cell scenario with a radius of 500 m. We assume that each CUE  has been assigned a RB and a RB can be allocated to multiple D2D pairs. So we set the number of RBs to be the same as the number of CUEs. The size of experience replay buffer is set to 1000000. The CUEs and D2D pairs are distributed randomly in a cell, where the communication distance of each D2D pair cannot exceed a given maximum distance 30 m. The detail parameters can be found in Table \ref{parameters}. The actor network in our proposed frameworks is a four-layer fully connected neural network with two hidden layers. The numbers of neurons in the two hidden layers are 512 and 128, respectively. The critic network in our proposed frameworks is a five-layer fully connected neural network with three hidden layers. The numbers of neurons in the three hidden layers are 1024, 512 and 256, respectively. Relu function is used as the activation function. The learning rates of actor and critic parts are 0.0001 and 0.001, respectively. The reward discount factor $ \gamma = 0.95 $. The UE noise figure is taken 8 dB. And the negative reward $ r_{neg} = -1 $. The channel model is set according to 3GPP Technical Specification \cite{access2010further}. In the first 2000 time slots of the system simulation, we use the random allocation method to allocate RBs to users, let the framework collect a certain number of samples for training, and then apply our algorithm for spectrum allocation. All simulations were conducted on Pytorch deep learning framework with a NVIDIA TESLA M40 GPU, 24 G memory size.
\begin{table}[!t]
\caption{Simulation Parameters}
\begin{center}
\begin{tabular}{|c|c|}
	\hline
	\textbf{Parameter}& \textbf{Value} \\
	\hline
	Cell radius & 500 m \\
	\hline
	Maximum D2D pair distance & 30 m \\
	\hline
	Carrier frequency & 2 GHz\\
	\hline
	RB bandwidth & 180 KHz\\
	\hline
	Number of CUEs & 10\\
	\hline
	Number of RBs & 10\\
	\hline
	Number of D2D pairs & 10, 20, ..., 50\\
	\hline
	BS transmission power ($ P^{b} $) & 46 dBm\\
	\hline
	D2D transmission power ($ P^{d} $) & 13 dBm\\
	\hline
	Cellular link pathloss & $ 128.1+37.6 \log _{10} (d[km]) $\\
	\hline
	D2D link path loss exponent & 4\\
	\hline
	UE thermal noise density & -174 dBm/Hz\\
	\hline
	CUE target SINR threshold ($ \xi^{c}_{min} $) & 0 dB\\
	\hline
	UE noise figure & 8 dB \\
	\hline
	Negative reward ($ r_{neg} $) & -1 \\
	\hline
\end{tabular}
\label{parameters}
\end{center}
\end{table}

\subsection{Simulations Results}
\begin{figure}[!t]
\centering
\includegraphics[width=3.5in]{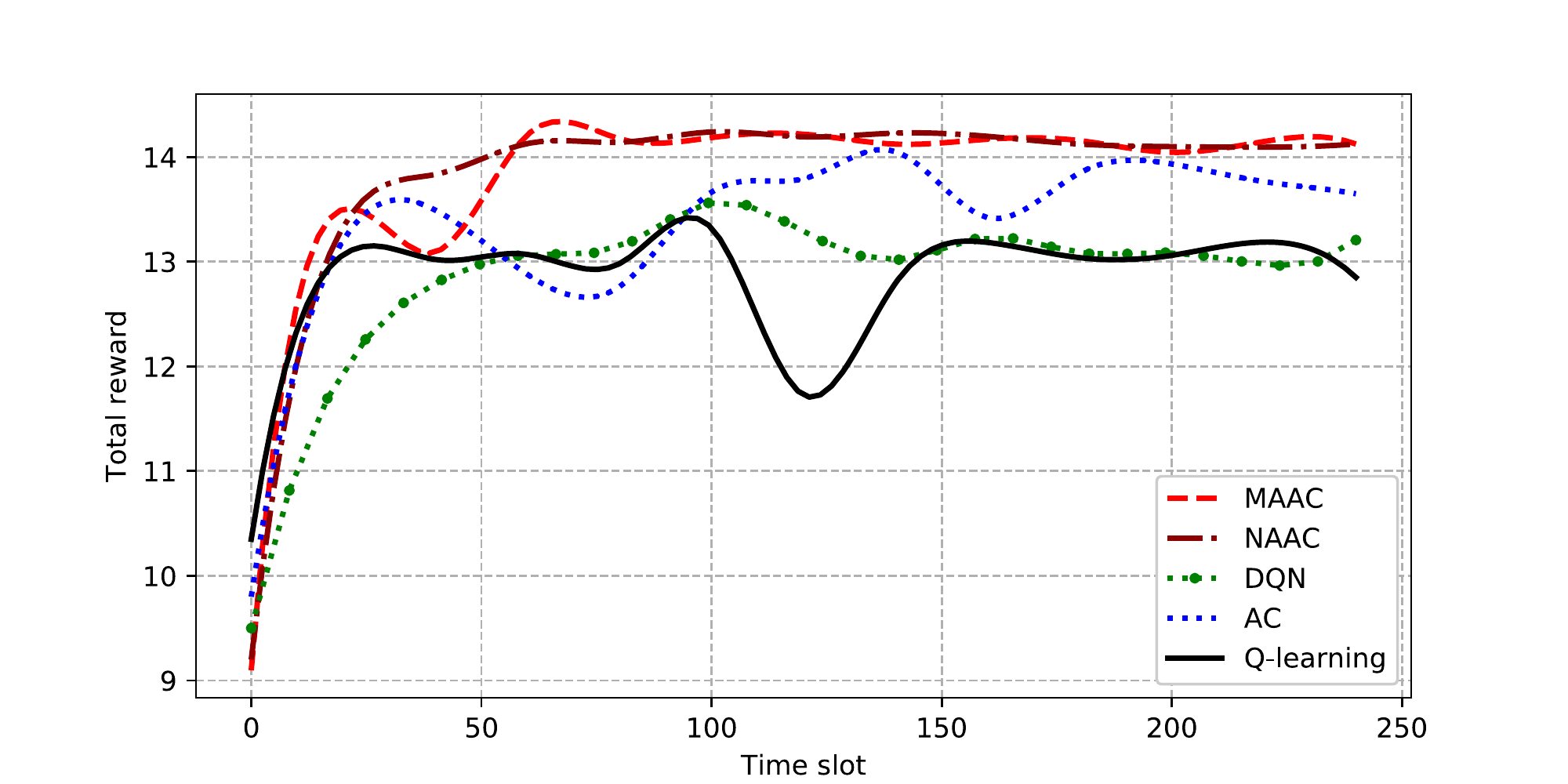}
\caption{Comparison of convergence performance during training process.}
\label{reward}
\end{figure}
Fig. \ref{reward} compares the convergence of the five approaches in terms of the total reward performance when the number of D2D pairs is 10 and NAAC works with the number of neighbor users $ \lambda=3 $. Total reward is the sum of the rewards obtained by the agents corresponding to all D2D pairs. Since SLA is an online learning algorithm that does not have an offline training process. From Fig. \ref{reward}, the proposed MAAC and NAAC converges to the maximum total reward with only 60 time slots. The proposed two methods achieve the larger total reward performance while the convergence is most stable (less fluctuations) compared to the other three algorithms. The total reward performance and convergence of Q-learning are the worst since Q-learning does not work well when the state-action space is vary large. DQN solves the mapping problem of high-dimensional space by introducing a DNN to approximate the complex mapping between state-action space. Compared to Q-learning, DQN improves in both total reward performance and convergence. The performance of AC algorithm is better than Q-learning and DQN since it optimizes the policy by combining the process of the policy learning and value learning with good convergence properties. However, none of the above three algorithms considers the impact of multi-agent environment on stability of training process and the cooperation between multiple agents (D2D pairs) on system performance. The two proposed approaches introduce the state and action information of extra D2D pairs to assist the training process, greatly improving the stability of the training process, and achieving a higher total reward performance and converging quickly.

\begin{figure}[!t]
\centering
\includegraphics[width=3.5in]{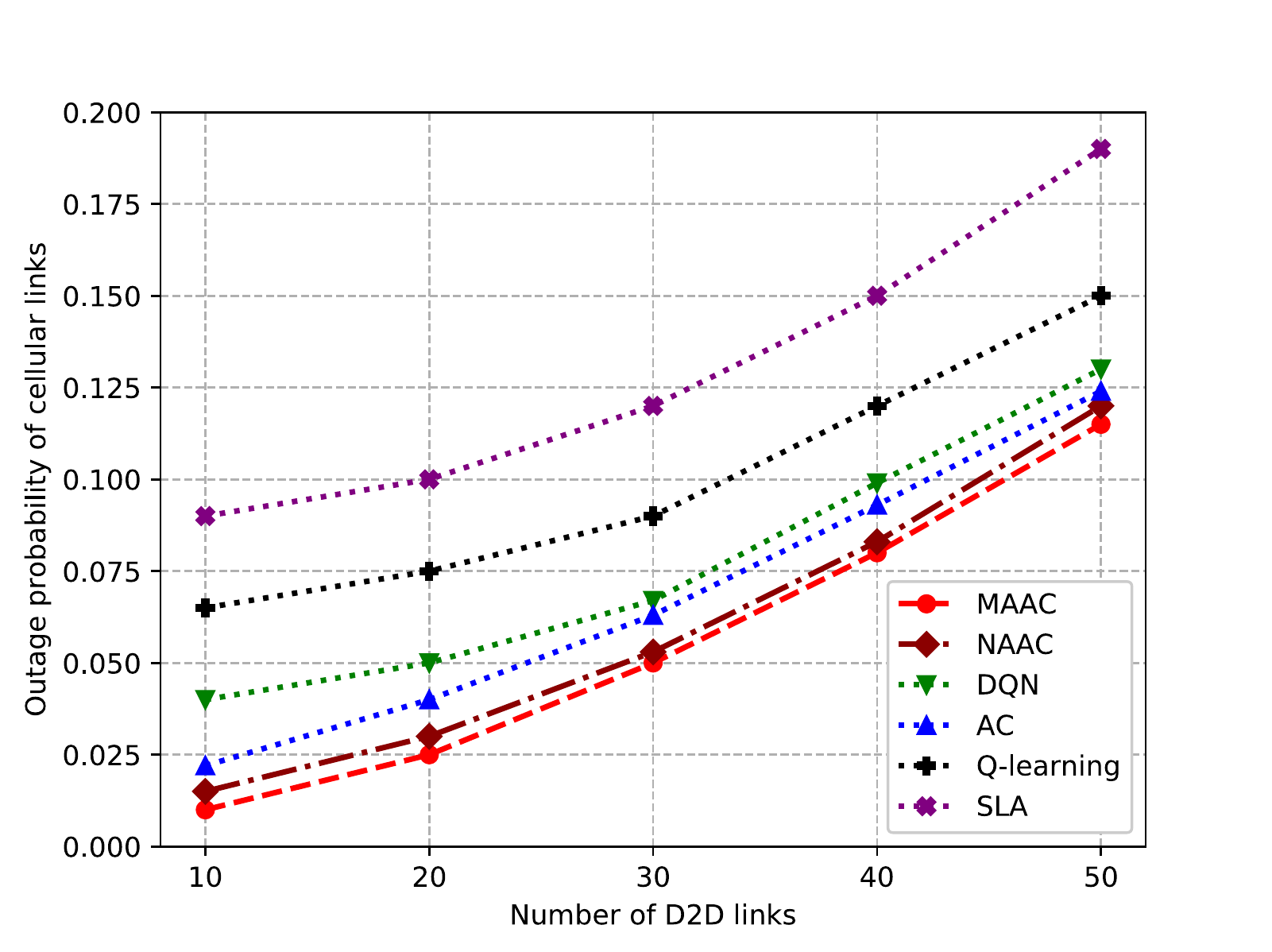}
\caption{Outage probability of cellular links versus the number of D2D links.}
\label{cue_op}
\end{figure}
The outage probability can reflect the reliability of the communication links. In Fig. \ref{cue_op}, we show the outage probability of cellular links as a function of the number of D2D pairs $ N $ and NAAC works with $ \lambda=3 $. The outage probability of cellular links increases as the number of D2D links grows since there are more D2D pairs sharing the spectrum with the CUEs, which causes the CUEs to suffer more severe cross-layer interference. The two proposed methods are better than the other four algorithms because the reward function in the proposed frameworks penalizes the policy that does not meet the SINR threshold of CUEsc and the frameworks introduce the states and actions of extra D2D pairs to assist the training process. Therefore, the policies between different D2D pairs can be coordinated with each other to prevent multiple D2D pairs from simultaneously selecting the same RB to cause severe cumulative interference to the CUE. From Fig. \ref{cue_op}, the MAAC algorithm achieves the lowest outage probability, which is 0.005 lower than the NAAC algorithm. Since the MAAC algorithm uses the information of all D2D pairs for centralized training, the learned strategy more strictly meets the SINR constraints of CUEs than the NAAC that uses part of the D2D pairs' information for training.

\begin{figure}[!t]
\centering
\includegraphics[width=3.5in]{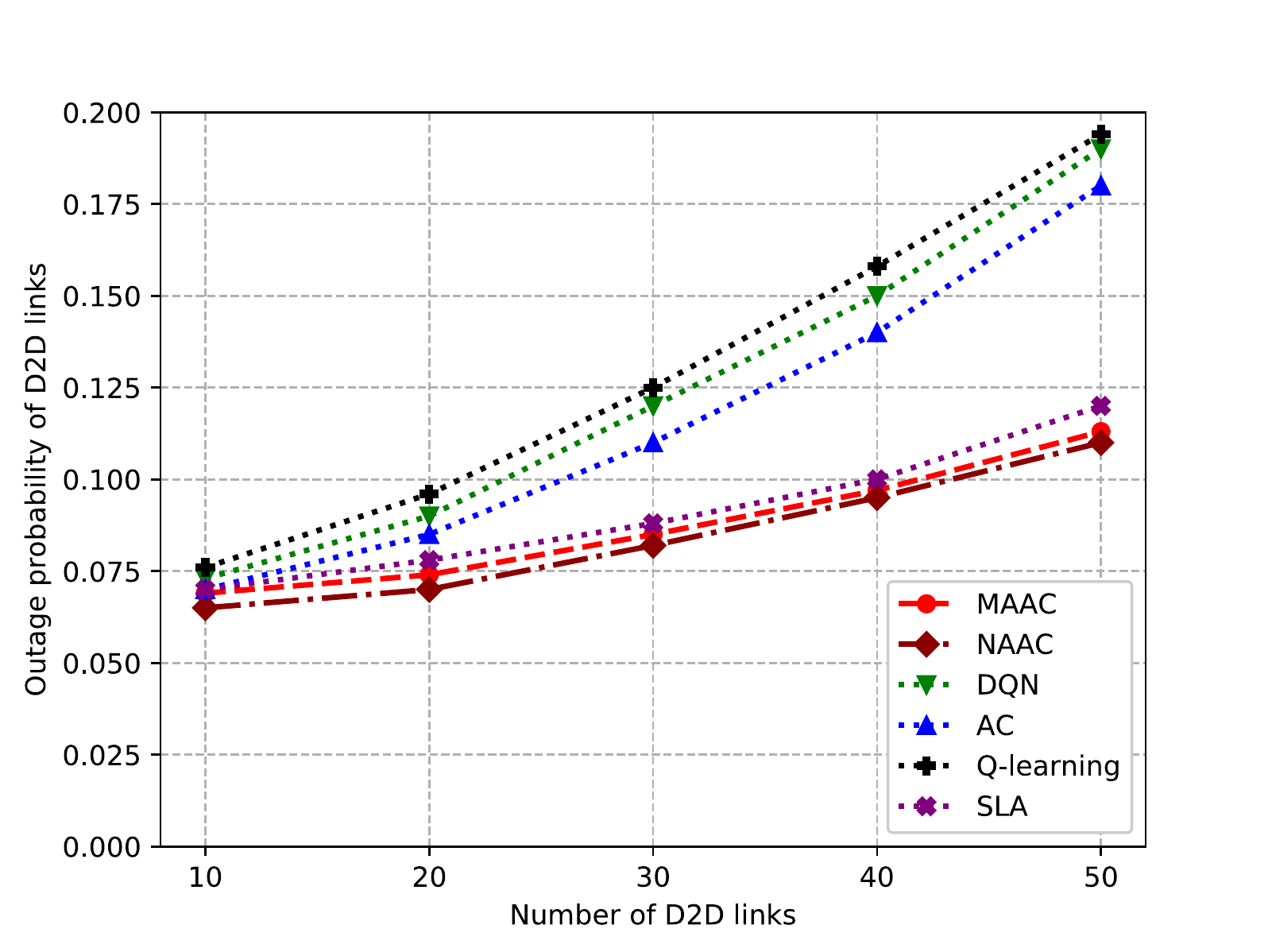}
\caption{Outage probability of D2D links versus the number of D2D links.}
\label{d2d_op}
\end{figure}
Fig. \ref{d2d_op} illustrates the outage probability of D2D links as a function of $ N $, where $ \lambda=3 $ for NAAC. The outage probability of D2D links increases as the number of D2D links grows since there are more D2D pairs sharing the spectrum and there will be more serious co-layer interference among them. From Fig. \ref{d2d_op}, the two proposed algorithms are obviously superior to other algorithms since our proposed algorithms make full use of the cooperation between D2D pairs so that the policy learned by each agent can coordinate with each other and avoid selecting the same RB at the same time, which leads to better transmission quality. For the Q-learning, DQN and AC, the outage probabilities increase significantly with the number of D2D links. This is because the policies learned by these three algorithms consider no information of other D2D pairs when they are executed, result in multiple D2D pairs to compete for the same RB, and seriously affect the transmission quality of D2D links. The SLA algorithm achieves performance close to our proposed algorithms since it estimates the interference experienced by the D2D pairs and takes an action based on this estimate.

\begin{figure}[!t]
\centering
\includegraphics[width=3.5in]{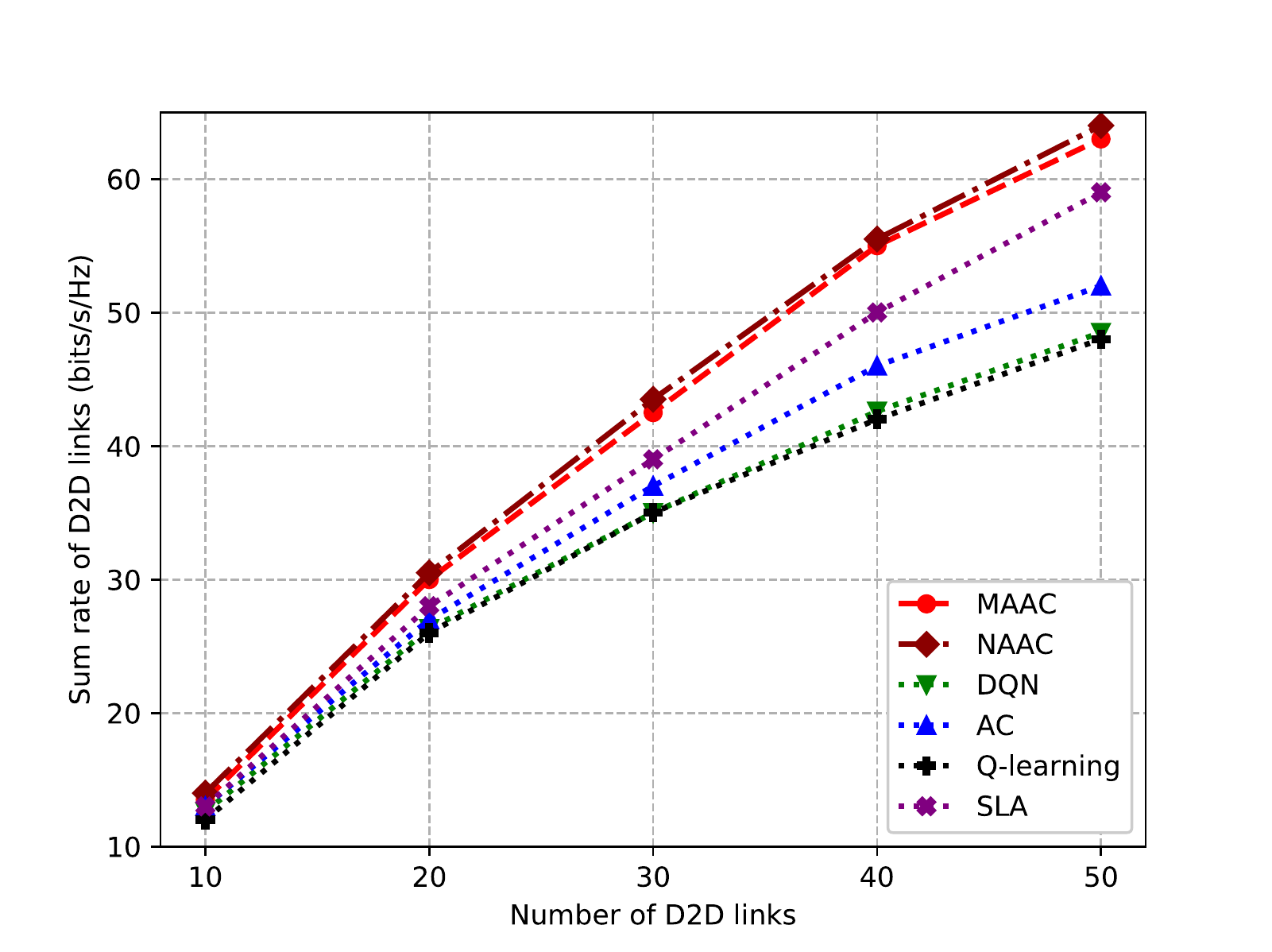}
\caption{Sum rate of D2D links versus the number of D2D links.}
\label{d2d_sum_rate}
\end{figure}
Fig. \ref{d2d_sum_rate} shows the the sum rate of D2D links as a function of $ N $, where $ \lambda=3 $ for NAAC. The the sum rate of D2D links increases as the number of D2D links grows since more D2D pairs are allocated to RBs. When the number of D2D links increases, outage probability increases due to higher interference. Therefore, the slope of all the curves in Fig. \ref{d2d_sum_rate} is decreasing.  The proposed methods are significantly better than the other four algorithms and the advantages become more significant as the number of D2D links increases. Since the other four distributed algorithms can only achieve individual optimization, the effect of global optimization cannot be guaranteed, but the proposed methods adopt a framework of centralized training with decentralized execution, which can optimize the sum rate of D2D links.  The performance indicators in Fig. \ref{d2d_op} and Fig. \ref{d2d_sum_rate} are indicators related to D2D communication. NAAC is slightly better than MAAC in performances related to D2D  communication. That is because our optimization goal is to maximize the sum rate of D2D  communications while ensuring the outage probability of cellular links. In order to achieve  this goal, our reinforcement learning framework penalizes actions that fail to meet the SINR  requirements of cellular users. MAAC uses more information to better achieve its goals, so it achieves the lowest outage probability for cellular users. Ensuring the communication quality of cellular users is bound to lose the performances of D2D users, so the performances of MAAC is	slightly worse to NAAC in Fig. \ref{d2d_op} and Fig. \ref{d2d_sum_rate}.

\begin{figure}[!t]
\centering
\includegraphics[width=3.5in]{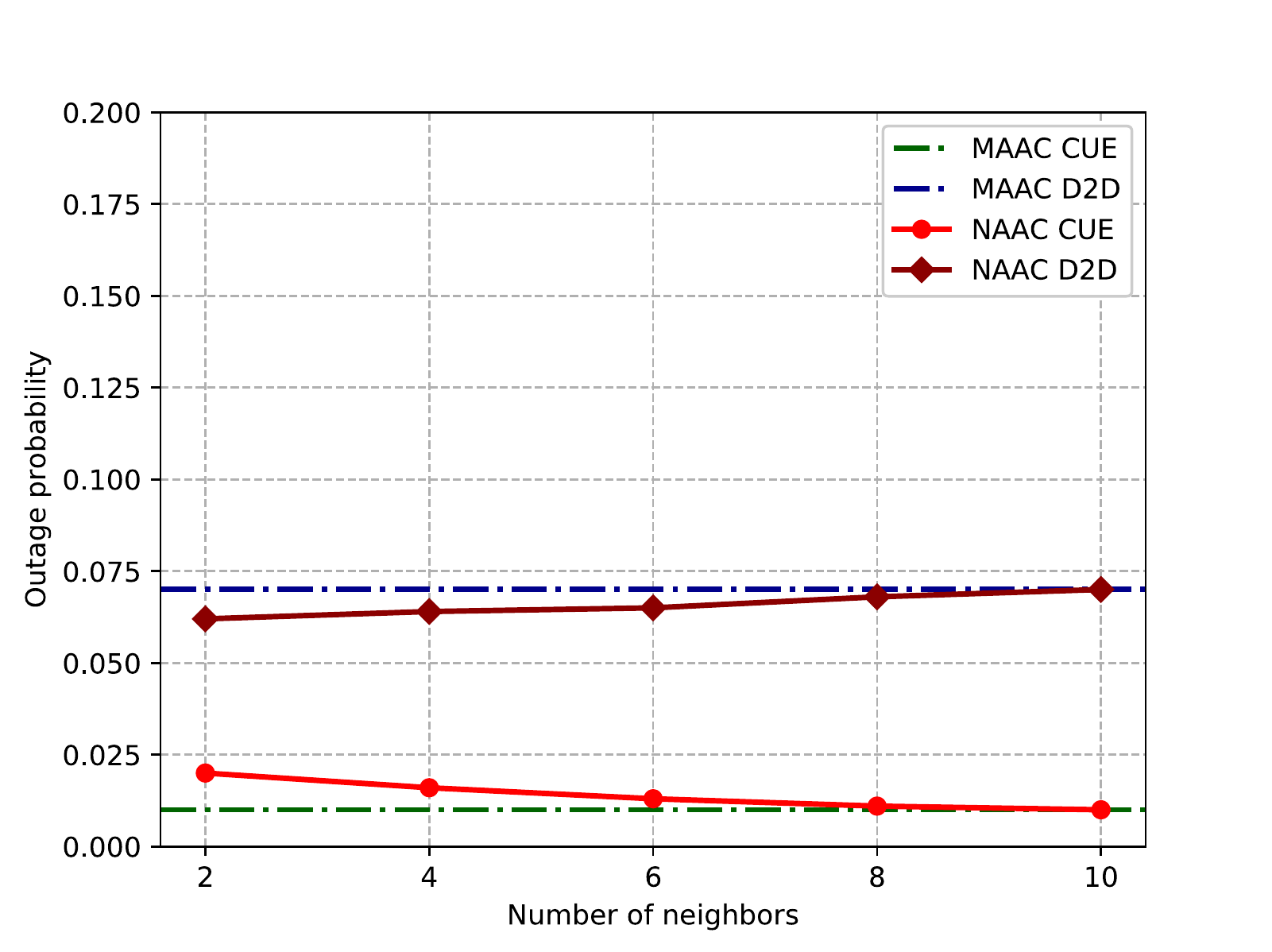}
\caption{Outage probability of cellular and D2D links versus the number of neighbor users.}
\label{cue_d2d_op}
\end{figure}
In Fig. \ref{cue_d2d_op}, we show the outage probability of cellular and D2D links as a function of the number of neighbor users $ \lambda $ in NAAC with number of D2D pairs $ N=10 $. For the NAAC, the outage probability of the cellular links is greater than the MAAC when the $ \lambda $ is small, and the outage probability decreases continuously with the increase of $ \lambda $ until it is equal to the MAAC. Since the NAAC obtains more comprehensive information during training with the increase of $ \lambda $, the trained policy can provides more reliable protection for the transmission quality of CUEs. In addition, the outage probability of D2D links with the NAAC is smaller than the MAAC when the $ \lambda $ is small, and the outage probability increases with $ \lambda $ until it is equal to MAAC. The reason is that cellular communications have a higher priority than D2D communications, the policy trained by the MAAC algorithm using global information sacrifices some D2D links transmission quality to meet the transmission quality requirements of cellular users.

\begin{figure}[!t]
\centering
\includegraphics[width=3.5in]{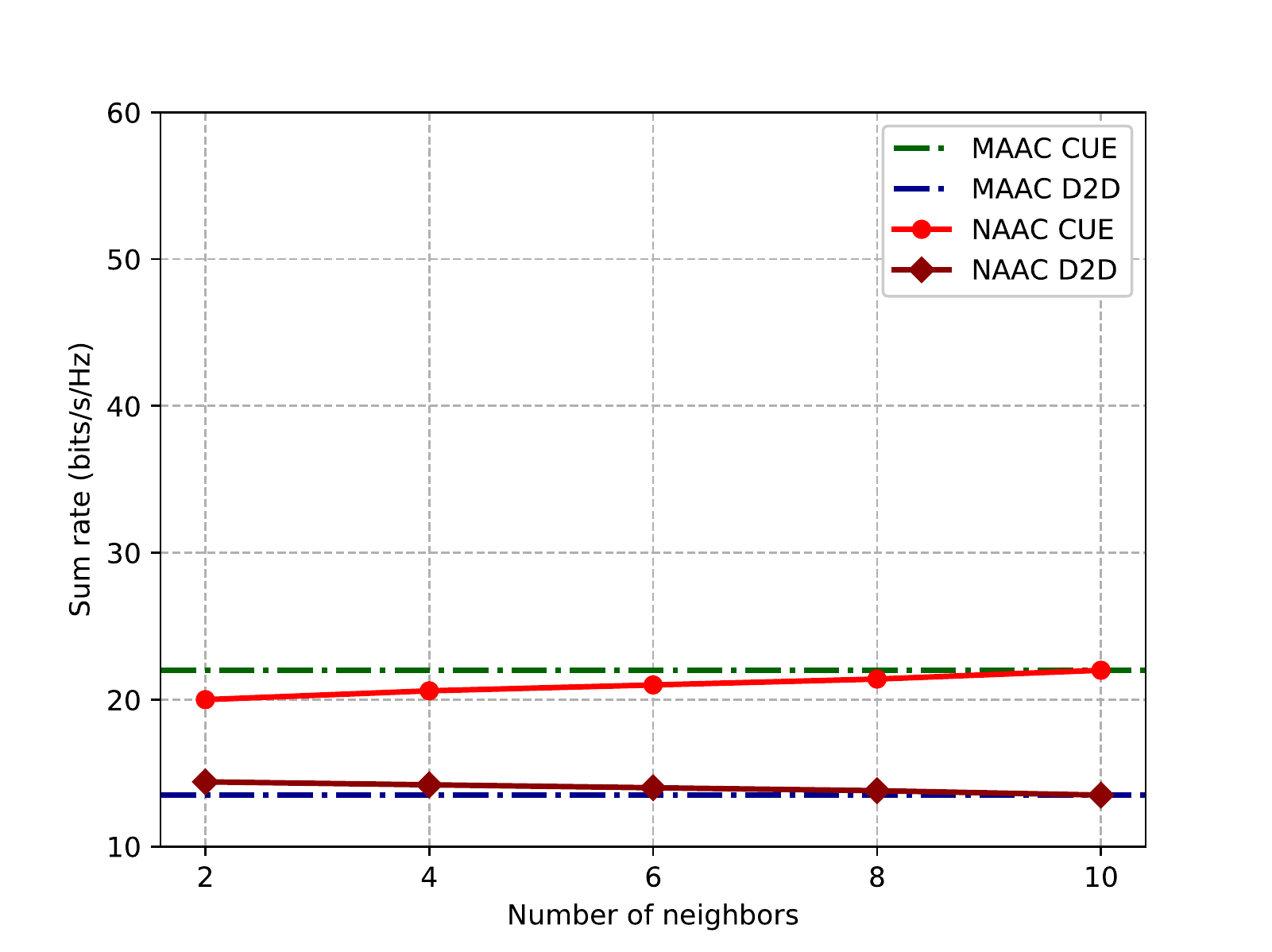}
\caption{Sum rate of cellular and D2D links versus the number of neighbor users.}
\label{cue_d2d_sum_rate}
\end{figure}
Fig. \ref{cue_d2d_sum_rate} illustrates the sum rate of cellular and D2D links as a function of $ \lambda $ in NAAC with $ N=10 $. For the NAAC, the sum rate of cellular links is 2 bit/s/Hz lower than the MAAC at $ \lambda=2 $, and the rate increases as $ \lambda $. Since the NAAC uses part of the information of the neighbor D2D pairs for training, and the trained policy does not adequately guarantee the transmission quality of the CUEs compared to the MAAC. In addition, the sum rate of D2D links with the NAAC is larger than the MAAC when $ \lambda $ is small, and the sum rate continues to decrease with the increase of $ \lambda $ until it is equal to the MAAC. The reason is that the constraint of satisfying the SINR of the cellular user has higher priority than increasing the D2D rate. As the $ \lambda $ increases, the information obtained by the NAAC algorithm during training is more comprehensive, and the trained policy satisfies the constraint more strictly, resulting in a loss of a portion of the D2D sum rate. According to the simulation results in Fig. \ref{cue_d2d_op} and Fig. \ref{cue_d2d_sum_rate}, when using the NAAC, $ \lambda $ can be flexibly adjusted according to different communication scenarios to meet the various communication requirements.

\subsection{Discussion}
The proposed two frameworks exploits the advantages of the centralized and distributed schemes. Compared with the centralized methods, our methods are executed without requiring the global information, which significantly reduces the signaling overhead and alleviates the computational pressure of the BS. Compared with distributed method, our methods use the historical information of the extra users to learn the policies of mutual cooperation, avoiding frequent real-time information exchange between users, more suitable for user-intensive communication scenario. In addition, our methods can transfer complex training processes to the cloud (BS), significantly reduces the computing complexity of algorithm execution. The two proposed methods have their own advantages respectively. MAAC uses the historical information of all users to assist in training. The trained policies can meet very strict transmission quality requirements and are suitable for high-reliability wireless communication scenarios. NAAC uses a fixed number of users' historical information to assist training, has better generalization ability, can be scale well to a larger network and has lower training complexity.

\section{Conclusion}
This paper has studied the resource management problem in D2D underlay communications and formulated the intelligent spectrum allocation problem as a decentralized multi-agent deep RL model to improve the sum rate of D2D links while ensuring the transmission quality of CUEs. In order to make full use of the performance gains brought by cooperation between users, the MAAC framework of centralized training with distributed execution is adopted, which not only requires no signaling interaction but also ensures the convergence of the algorithm. In addition, the NAAC framework with lower computing complexity and better generalization ability is proposed. The simulation results show that the proposed approaches can effectively guarantee the transmission quality of the CUEs and greatly improve the sum rate of D2D links as well as have better convergence, compared with other existing approaches. The proposed methods can be used to address the intelligent resource management problem in a D2D-based Internet of Vehicle networks. In the future work, we plan  to combine the proposed approaches with continuous-valued power control, and  design an integrated deep reinforcement learning framework that automatically selects RB and transmit power to further improve the effectiveness and robustness of the algorithm.

% or



%\section{}

%\section*{Acknowledgment}


\ifCLASSOPTIONcaptionsoff
\newpage
\fi



%\bibliographystyle{IEEEtran}

\bibliographystyle{IEEEtran}
\footnotesize
\bibliography{IEEEabrv,ref}

\end{document}